\documentstyle[11pt,epsf]{article}
\def\usq{U_{12}^2}
\def\vsq{V_{12}^2}
\def\sig{\sigma}
\def\xup{\sig_U^+}
\def\xum{\sig_U^-}
\def\xpp{\sig_P^+}
\def\xpm{\sig_P^-}
\def\ebar{\anti E}
\def\ehat{\what E}
\def\kp{\vec p_{\pi^+}}
\def\km{\vec p_{\pi^-}}
\def\Em{E_{\pi^-}}
\def\Ep{E_{\pi^+}}
\def\pn{\vec p_{\cnone}}
\def\pc{\vec p_{\cmone}} 
\def\pcp{\vec p_{\cpone}}
\def\pnp{\vec p_{\anti{\cnone}}}
\def\crs{\km\times\kp}
\def\dt{\km\cdot\kp}
\def\pep{p_{e^+}}
\def\pem{p_{e^-}}
\def\pg{p_\gam}
\def\qll{Q_{LL}}
\def\qrr{Q_{RR}}
\def\qlr{Q_{LR}}
\def\qrl{Q_{RL}}

\def\lsim{\mathrel{\raise.3ex\hbox{$<$\kern-.75em\lower1ex\hbox{$\sim$}}}}
\def\gsim{\mathrel{\raise.3ex\hbox{$>$\kern-.75em\lower1ex\hbox{$\sim$}}}}

\def\square{\boxxit{0.4pt}{\fillboxx{7pt}{7pt}}\hskip-0.4pt}
    \def\boxxit#1#2{\vbox{\hrule height #1 \hbox {\vrule width #1
             \vbox{#2}\vrule width #1 }\hrule height #1 } }
    \def\fillboxx#1#2{\hbox to #1{\vbox to #2{\vfil}\hfil}    }

\def\ibid{{\it ibid.}}

\def\etal{{\it et al.}}

\def\cw{c_W}

\def\sb{s_\beta}
\def\cb{c_\beta}
\def\dmax{D_{\rm max}}
\def\qks{Q_{KS}}

\def\epipr{E_\pi^*}

\def\mpi{m_{\pi}}

\def\cpmtwo{\wt \chi^{\pm}_2}

\def\dmchi{\Delta m_{\tilde\chi}}

\def\delgs{\delta_{GS}}

\def\emiss{~{/ \hskip-8pt E}}

\def\mslash{~{/ \hskip-9pt M}}
\def\pslash{~{/ \hskip-6pt p}}
\def\ptslash{~{/ \hskip-6pt p}_T}
\def\mchichi{M_{\chi\chi}}

\def\eg{{\it e.g.}}

\def\tanb{\tan\beta}

\def\mz{m_Z}
\def\mw{m_W}

\def\wp{W^+}
\def\wm{W^-}

\def\wmp{W^{\mp}}

\def\cnone{\wt\chi^0_1}

\def\cntwo{\wt\chi^0_2}
\def\cnthree{\wt\chi^0_3}
\def\cnfour{\wt\chi^0_4}
\def\snu{\wt\nu}

\def\snue{\wt\nu_e}

\def\msnu{m_{\snu}}
\def\mcnone{m_{\cnone}}
\def\mcntwo{m_{\cntwo}}

\def\wt{\widetilde}

\def\cpone{\wt \chi^+_1}
\def\cmone{\wt \chi^-_1}
\def\cpmone{\wt \chi^{\pm}_1}
\def\cmpone{\wt \chi^{\mp}_1}

\def\mcpmone{m_{\cpmone}}

\def\cpmtwo{\wt \chi^{\pm}_2}

\def\wtil{\widetilde}
\def\what{\widehat}

\def\gam{\gamma}

\def\anti{\overline}
\def\epem{e^+e^-}

\def\rts{\sqrt s}

\def\eg{{\it e.g.}}

\def\anti{\overline}
\def\wp{W^+}
\def\wm{W^-}
\def\mw{m_W}
\def\mz{m_Z}

\def\fbi{~{\rm fb}^{-1}}
\def\fb{~{\rm fb}}

\def\abi{~{\rm ab}^{-1}}
\def\mev{~{\rm MeV}}
\def\gev{~{\rm GeV}}
\def\tev{~{\rm TeV}}

\def\MPL #1 #2 #3 {{ Mod.~Phys.~Lett.}~{\bf#1} (#3) #2}
\def\NPB #1 #2 #3 {{ Nucl.~Phys.}~{\bf #1} (#3) #2}
\def\PLB #1 #2 #3 {{ Phys.~Lett.}~{\bf #1} (#3) #2}
\def\PR #1 #2 #3 {{ Phys.~Rep.}~{\bf#1} (#3) #2}
\def\PRD #1 #2 #3 {{ Phys.~Rev.}~{\bf #1} (#3) #2}
\def\PRL #1 #2 #3 {{ Phys.~Rev.~Lett.}~{\bf#1} (#3) #2}
\def\RMP #1 #2 #3 {{ Rev.~Mod.~Phys.}~{\bf#1} (#3) #2}
\def\ZPC #1 #2 #3 {{ Z.~Phys.}~{\bf #1} (#3) #2}
\def\IJMP #1 #2 #3 {{ Int.~J.~Mod.~Phys.}~{\bf#1} (#3) #2}
\def\NIM #1 #2 #3 {{ Nucl.~Inst.~and~Meth.}~{\bf#1} {#3} #2}
\def\JHEP #1 #2 #3 {{ JHEP}~{\bf#1} (#3) #2}

\newcommand{\nc}{\newcommand}
\nc{\beq}{\begin{equation}}   \nc{\eeq}{\end{equation}}
\nc{\bea}{\begin{eqnarray}}   \nc{\eea}{\end{eqnarray}}
\nc{\baa}{\begin{array}}      \nc{\eaa}{\end{array}}
\nc{\bit}{\begin{itemize}}    \nc{\eit}{\end{itemize}}
\nc{\ben}{\begin{enumerate}}  \nc{\een}{\end{enumerate}}
\nc{\bce}{\begin{center}}     \nc{\ece}{\end{center}}
\def\beqa{\begin{eqnarray}}
\def\eeqa{\end{eqnarray}}

\def\tightenlines{\def\baselinestretch{1.3}\small\normalsize}
\tightenlines
\begin{document}
\font\fortssbx=cmssbx10 scaled \magstep2
\hbox to \hsize{
$\vcenter{
\hbox{\fortssbx University of California - Davis}
}$
\hfill
$\vcenter{\normalsize
\hbox{\bf UCD-2001-3} 
\hbox{\bf hep-ph/0103167}
\hbox{March, 2001}
}$
}
\begin{center}
{ \Large \bf  Probing Models with
Near Degeneracy of the Chargino and LSP at a Linear {\boldmath$\epem$} Collider}
\rm
\vskip1pc
{\large\bf John F. Gunion and Stephen Mrenna}\\
\medskip
{\it Davis Institute for High Energy Physics}\\
{\it University of California, Davis, CA 95616}\\
\end{center}

\begin{abstract}
For some choices of soft SUSY--breaking parameters, the LSP is a stable
neutralino $\cnone$, the NLSP is a chargino $\cpmone$ almost degenerate in
mass with the LSP ($\dmchi\equiv\mcpmone-\mcnone\sim \mpi-$few GeV),
and all other sparticles are relatively heavy.  We discuss
the potential of a $\rts\sim 600\gev$ $\epem$ collider 
for studying such models.
\end{abstract}

\noindent{\bf 1. Introduction.} 
As part of the process of planning for future HEP experimental facilities,
it is important to evaluate as many motivated scenarios for new
physics as possible.  Certainly, supersymmetry ranks as one
of the most successful models of physics beyond the Standard
Model, since it can approximately reproduce Standard Model
predictions at low energies, while explaining the hierarchy
problem.  However, the mechanism of supersymmetry breaking is
not well understood. In general, the different sources of breaking
-- gravitational interactions, gauge interactions,
the conformal anomaly, etc. -- lead to different hierarchies of
sparticle masses.  Many sources may
be present at once, so the true model may be quite complicated.
Here, we explore a relatively well-motivated set of models in which
the gaugino masses are non-universal at the GUT scale and, in particular,
are such that there is a quite small mass splitting between the lightest
chargino ($\cpmone$) and the lightest neutralino ($\cnone$), which is the LSP.
Small $\dmchi\equiv \mcpmone-\mcnone$ occurs
when $ M_2<M_1 \ll |\mu|$, which arises 
naturally when the gaugino masses are dominated by or entirely
generated by loop corrections. Models of this type include
the O--II superstring model \cite{ibanez,cdg1,cdg2,cdg3} 
and the (AMSB) models in which SUSY--breaking
arises entirely from the conformal anomaly 
\cite{Randall:1999uk,Giudice:1998xp,Pomarol:1999ie}.
The same hierarchy also occurs when SUSY is broken by
an $F$--term that is not an SU(5) singlet but
rather is a member of the ${\bf 200}$ representation
contained in $({\bf 24}{\bf \times} 
{\bf 24})_{\rm symmetric}={\bf 1}\oplus {\bf 24} \oplus {\bf 75}
 \oplus {\bf 200}$ \cite{sm96nonuniv}.
Techniques for detecting and studying supersymmetry 
are very dependent upon $\dmchi$ and on the relative magnitude
of $\mcpmone\sim\mcnone$ as compared to other supersymmetric particles.
Previous discussions of small $\dmchi$ scenarios at lepton colliders
include early studies in the
context of the O-II model \cite{cdg1,cdg2,cdg3} and
various studies specific to the AMSB boundary conditions
\cite{Gherghetta:1999sw,Feng:1999fu,Gunion:2000jr,Paige:1999ui}.
This study will focus on signals and parameter determination for small $\dmchi$
at a $\rts\sim 450-600\gev$, $L\sim 1\abi$ linear $\epem$ collider (LC)
when the only SUSY
particles kinematically accessible at the LC are the $\cnone$
and $\cpmone$.

To illustrate the possibilities for $\dmchi$ and other gaugino masses,
it is useful to give some tree-level 
results for the gaugino mass parameters
at the scale $\mz$. The O--II
model with $\delgs=-4$ yields $M_3:M_2:M_1=6:10:10.6$,
the O--II model with $\delgs=0$ (equivalent to the 
simplest version of the conformal anomaly, AMSB, approach) yields
$M_3:M_2:M_1=3:0.3:1$, 
while the ${\bf 200}$ model yields $6:4:10$. As a result:
in the ${\bf 200}$  
 model and the O--II $\delgs=0$  (or pure AMSB) model,
$M_2\ll M_1$ and $\dmchi\equiv \mcpmone- \mcnone$
can be smaller than $\sim m_\pi$ at tree-level;
in the O--II $\delgs=-4$ case, $M_2$ is only slightly
less than $M_1$ implying that $\dmchi<$ a few GeV is very typical ---
$\dmchi<1\gev$ if $|\mu|\gsim 1\tev$ as 
when RGE electroweak symmetry breaking is imposed \cite{cdg2}.
Loop corrections can be significant.  In particular, even if
$\dmchi<\mpi$ at tree-level,
radiative corrections \cite{Gherghetta:1999sw,radcor,matchev} 
usually increase $\dmchi$ to above $m_\pi$
(see e.g. the $\dmchi$ graphs in \cite{cdg2});
$\dmchi<\mpi$ is only possible for very special parameter choices. 
Most typically, $\dmchi$ is predicted to lie 
in the range from slightly above $\mpi$ to several GeV.

All other SUSY particles could have masses substantially above $\mcpmone$.
In particular, in the O--II $\delgs=0$ and AMSB models, 
$\mcntwo$ is significantly larger than $\mcpmone$.
Only in the O--II $\delgs=-4$ model is 
$\mcntwo$ close to $\mcpmone$.  
In most cases, large $|\mu|$ is required by radiative symmetry breaking
so that the $\cpmtwo$, $\cnthree$ and $\cnfour$ are very heavy.
The masses for the squarks and sleptons are uncertain, but
could be quite large, which is the case on which we focus. In particular,
we neglect sneutrino exchange contributions to
the $\cpone\cmone$ and $\gam\cpone\cmone$ cross sections. For example,
both cross sections start to become 
increasingly suppressed for $\rts\sim 600\gev$
as the electron sneutrino mass is decreased below 1 TeV; see, for example,
\cite{Datta:1999yw}.

The neutralino and chargino couplings to $W$ and $Z$ bosons
in the wino-LSP scenario were reviewed in \cite{Gunion:2000jr}.
One finds that the $Z\cnone\cnone$, $Z\cnone\cntwo$,
$Z\cntwo\cntwo$, and 
$\wmp\cpmone\cntwo$ couplings (and corresponding cross sections)
 are all small, while
$\epem\to Z,\gam\to\cpone\cmone,\gam\cpone\cmone$ 
and $\epem\to \wmp \cpmone\cnone$ 
can have large rates.

\begin{table}[ht]
\renewcommand{\arraystretch}{0.7}
  \begin{center}
    \begin{tabular}[c]{|c|c|c|c|c|c|c|c|} \hline
$\dmchi(\mev)$ & 125 & 130 & 135 & 138 & 140 & 142.5 & 150 \\
$c\tau(\mbox{cm})$ & 1155 & 918.4 & 754.1 & 671.5 & 317.2 & 23.97 & 10.89 \\
\hline
$\dmchi(\mev)$ & 160 & 180 & 185 & 200 & 250 & 300 & 500 \\
$c\tau(\mbox{cm})$ & 6.865 & 3.719 & 3.291 & 2.381 & 1.042 & 0.5561 & 0.1023 \\
\hline
$\dmchi(\mev)$ & 600 & 700 & 800 & 900 & 1000 & 1500 & 2000 \\
$c\tau(\mbox{cm})$ & 0.055 & 0.033 & 0.019 & 0.011 & 0.0072 & 0.0013 & 0.00036 \\
\hline
    \end{tabular}
    \caption{Summary of $c\tau$ values as a function of $\dmchi$.}
    \label{ctaus}
  \end{center}
\end{table}

The most critical ingredients in the phenomenology of such models
are the lifetime and the decay modes of the $\cpmone$, 
which in turn depend almost entirely
on $\dmchi$ when the latter is small. The $c\tau$ and branching ratios
of the $\cpmone$ as a function of $\dmchi$ have been computed
in \cite{cdg3}.
For $\dmchi<\mpi$, only $\cpmone\to e^{\pm}\nu_e\cnone$ is important and
$c\tau>10$~m. Once $\dmchi>\mpi$, the $\cpmone\to \pi^\pm\cnone$ mode
turns on and is dominant for $\dmchi\lsim 800\mev$, at which
point the multi--pion modes start to become important:
correspondingly, one finds $c\tau\lsim 10-20$~cm
for $\dmchi$ just above $\mpi$ decreasing to $c\tau < 100~\mu$m
by $\dmchi\sim 1\gev$. 
For later reference, we give some
specific values of $c\tau$ as a function of $\dmchi$ in Table~\ref{ctaus}.

Finally, we wish to comment on what values for $\mcpmone$ are most natural
in the context we are considering.  The constraint from fine-tuning,
though somewhat ill-defined, does provide some guidance.
The degree of fine-tuning is largely controlled by the magnitude of the
gluino mass.  In the models discussed above there are considerable
differences in the ratio of $M_3/M_2$.  This ratio is largest ($\sim 9$)
in the ($\delgs=0$)/AMSB scenario and fine tuning increases 
rapidly \cite{Bastero-Gil:2000gu}
with $M_2\sim \mcpmone$  to levels that are
highly problematical once $\mcpmone>200\gev$. In the other scenarios
$M_3$ is close to $M_2$ and fine-tuning is a less severe concern.
The main focus of this paper will be on scenarios in which
the chargino and neutralino are highly degenerate, which
is most naturally the case in the ($\delgs=0$)/AMSB scenario.  Thus,
our focus will be on procedures relevant for $\mcpmone\lsim 200\gev$
and $\dmchi<1\gev$.

\noindent{\bf 2. General Discussion of Detector and Signals.}
For our discussions, we consider a detector with the components listed
in Table~\ref{detector}. (See \cite{Gunion:2000jr} for further details.)
The SVX, CT and PS all give (independent)
measurements of the $dE/dx$ from ionization of a track passing through them.
This makes it possible to distinguish a heavily--ionizing chargino
(which would be $\geq$ twice minimal ionizing [2MIP] for $\beta\gamma\leq
0.85$) from an isolated minimally ionizing particle [1MIP]. 
The net discrimination factor would probably be
of order ${\rm few}\times 10^{-5}$. In our simulations, we employed 
an efficiency of 90\% for tracks with $\beta\gamma<0.85$
\cite{glandsberg}.

The possible signals based upon detecting a non-promptly decaying
$\cpmone$ were detailed for a hadron collider in \cite{Gunion:2000jr}. 
Appropriately modified versions for $\cpone\cmone$
and $\gam\cpone\cmone$ production at the LC are listed in
Table~\ref{signals}.

\begin{table}[ht]
\renewcommand{\arraystretch}{0.7}
  \begin{center}
\begin{tabular}{|c|l|}
\hline
Component & Description \\
\hline
SVX & {Silicon vertex detector from close to beam pipe to $\sim$11 cm.}\\
CT  & {Central tracker starting just past SVX.}\\
PS  & {Pre--shower just outside the tracker.}\\
EC  & {Electromagnetic calorimeter.}\\
HC  & {Hadronic calorimeter.}\\
TOF & {Time--of--flight measurement after HC and 
just before MC.}\\
MC  & {Muon chamber with first layer after the HC and
just beyond the TOF.}\\
\hline
    \end{tabular}
    \caption{Summary of detector components referred to in the text.} 
    \label{detector}
  \end{center}
\end{table}

\begin{table}[p]
\renewcommand{\arraystretch}{0.9}
  \begin{center}
    \begin{tabular}[c]{|c|p{5.5in}|} \hline
Signal & Definition \\
\hline
LHIT &  At least one long, heavily--ionizing 
($\geq$ 2MIP's as measured by SVX+CT+PS),
large--$p_T$ track that reaches the MC.
The energy deposit in the HC in the track direction must be consistent
with expected ionization energy deposit
for the $\beta$ measured (using TOF and/or SVX+CT+PS), i.e.
no hadronic energy deposit.\\
\hline
TOF  & {At least one large--$p_T$ track seen in the SVX and CT along with a 
signal in the TOF delayed by 500 ps or more (vs. a particle with
$\beta=1$). HC energy deposit (in the direction of the track)
is required to be consistent with the ionization expected
for the measured $\beta$, i.e. no large hadronic deposit.}\\
\hline
DIT  & {At least one isolated, large--$p_T$ 
track in the SVX and CT that fails to reach the MC
and deposits energy in the HC no larger than that consistent with
ionization energy deposits for the measured (using SVX+CT+PS) $\beta$.
Heavy ionization in the SVX+CT+PS, corresponding
to $\beta<0.8$ or $\beta<0.6$ (DIT8 or DIT6), may be required.
Large $\emiss$.}\\
\hline
KINK & {At least one track that terminates in the CT, turning into a soft,
but visible,
charged--pion daughter--track at a substantial angle to parent.
Large $\emiss$.}\\
\hline
STUB & {At least one isolated, large--$p_T$ (as measured using SVX) track 
that registers in all SVX layers, but does not pass
all the way through the CT. 
Energy deposits in the EC and HC in the direction of the track
should be minimal. Large $\emiss$.}\\
\hline
SNT  & {One or more STUB tracks and large $\emiss$ with no additional trigger. 
Heavy ionization of the STUB in the SVX
corresponding to $\beta<0.8$ (SNT8) is required.}\\
\hline
S$\gam$ & {One or more STUB tracks and large $\emiss$ 
with a $p_T^\gam>10\gev$ trigger.
Heavy ionization of the STUB in the SVX
corresponding to $\beta<0.8$ (S$\gam$8) is required.}\\
\hline
HIP  & {At least one high--impact--parameter ($b\geq 5\sigma_b$)
soft pion track in the SVX, with $p_T^\gam>10\gev$ 
triggering and large $\emiss$, 
perhaps in association with a visible KINK in the SVX.}\\
\hline
$\gamma+\emiss$ & {Isolated, large--$E_T$ photon and large $\emiss$.
Relevant if the soft $\pi$'s cannot be detected.}\\
\hline
mSUGRA--like & jets or leptons + $\emiss$ \\
\hline
    \end{tabular}
    \caption{Summary of signals for $\cpone\cmone$ and
$\gam\cpone\cmone$ production. MIP refers to a minimally--ionizing--particle
such as a highly relativistic muon. For detector component notation, see
Table~\ref{detector}.}
    \label{signals}
  \end{center}
\end{table}

Assuming that only the $\cnone$ and $\cpmone$ are light,
SUSY particle production  will be primarily in the final states 
$\cpone\cmone$, $\gam\cpone\cmone$, and (when not
phase space limited) $\wmp\cnone\cpmone$.
(The $\gam\cnone\cnone$ cross section is suppressed due to 
the small $Z\cnone\cnone$ coupling.)
The possibilities for SUSY detection depend
largely upon which aspects (if any) of a $\cpmone$
in the final state are visible \cite{cdg1,cdg2,cdg3},
which in turn depends almost entirely on $\dmchi$.
We distinguish several interesting ranges for $\dmchi$:
\bit\itemsep=.1pt
\item
$\dmchi<m_\pi$:
the $\cpmone$ yields a `stable particle' LHIT 
and/or DIT track and is easily detected: $\cpone\cmone$ production
will be easily seen. 
\item
$m_\pi<\dmchi<1\gev$:
the $\cpmone\to \cnone\pi^\pm$ decay yields a soft $\pi$ track, 
possibly in association
with a STUB ($\dmchi<180\mev$) or HIP ($\dmchi<1\gev$) signature.
Direct 
$\cpone\cmone$ production then yields a $\emiss+\pi\pi$ final state,
where the $\emiss$ is associated with the $\cnone$'s.  
Since the $\pi$'s are very soft, backgrounds to this final state
from $\gam\gam$-induced interactions are very large, and it
is unlikely that the $\pi^+\pi^-$ SUSY signal can be isolated. 
One must tag $\cpone\cmone$ production using 
$\epem\to\gam\cpone\cmone$, leading to a $\gam+\emiss+\pi\pi$
final state, or employ the more kinematically
limited $\wmp\cnone\cpmone\to(\ell^\mp,q^\prime \,\anti q)+\emiss+\pi^\pm$ 
final states.
\item $1\gev<\dmchi<2\gev$: the $\cpmone$ decays with roughly equal
probability to $\cnone\pi^\pm$, $\cnone\pi^\pm\pi^0$ 
 and $\cnone\ell^\pm\nu$ ($\ell=e,\mu$).  The pion(s) or charged lepton will
be rather soft and a $\gam$ tag will probably still be necessary
to eliminate backgrounds. 
\item $\dmchi>2-3\gev$: the 
$\cpmone$  decays either to $\cnone+$multi-pion modes at the low end 
which start to resolve into jets at the higher end or to $\cnone\ell^\pm\nu$.
The visible decay products are sufficiently energetic that
$\gam\gam$ induced backgrounds can be rejected (using
a combination of event topology and $\emiss$) to the extent
necessary for mSUGRA-like mode detection of
direct $\epem\to\cpone\cmone$ production.
\eit
Sensitivity of LEP2 detectors to these various signatures
was sufficient to exclude \cite{delphi,l3}: (a) $\mcpmone< \rts/2$
in the `stable' and `standard' regions of $\dmchi$; and (b) $\mcpmone<80\gev$
(assuming the $\wtil\nu_e$ is heavy) 
in the $m_\pi\leq \dmchi\leq 2\gev$ region.
In particular, the backgrounds
to the `stable' and $\gamma$-tag+$\emiss$+soft-$\pi$ signals are very small.
In our analysis, we assume that they can continue to be neglected 
relative to our signal rates at the LC after simple cuts.
For the $\gam$-tag+soft-$\pi$ signal,  this may, at first sight, seem
problematical given that the signal cross section declines with increasing $s$,
whereas the two-photon ($\gam^*\gam^*$) 
collision backgrounds tend to increase logarithmically
with $s$. Below, we review the simple cuts that are likely to be very
effective in eliminating the two-photon backgrounds.

It is first important to understand the kinematics of the signal.
We will trigger on a photon with $p_T^\gam>10\gev$ and
$10^\circ<\theta_\gam<170^\circ$,
$10^\circ$ being the angle at
which  electromagnetic coverage by the detector is expected to begin. 
 We will require that 
$\mchichi^2=(p_{e^+}+p_{e^-}-p_\gam)^2$, the invariant mass-squared
of the $\cpone\cmone$ pair, be 
such that $\mchichi>2\mcpmone$ when searching for $\cpmone$'s of
a certain mass.  Of course, the search will start at $\mcpmone>80\gev$,
the LEP2 limit. Most of $\mchichi$ will be invisible,
being carried by the $\cnone\cnone$ pair. The $\pi^+$ and $\pi^-$ in
the final state will be very soft,
with energies basically set by the size of $\dmchi$,
and largely central, with some bias for the $\pi^+$ ($\pi^-$) to
move in the direction of the $e^-$ ($e^+$) beam. Thus, they will tend
to be somewhat back-to-back. However, their angle of acoplanarity 
will be very uniformly distributed. Finally, the $p_T^\gam$
of the trigger photon will be clustered at the lowest allowed values
and will be primarily balanced by the missing transverse momentum, $\ptslash$,
carried by the $\cnone$'s.

There are two different types of two-photon background
to the $\gam+\pi^+\pi^-+\emiss$ final state of interest. First, there are
$\epem\to\gam +\epem +\gam^*\gam^*\to\gam+\epem+X$, with $X=\pi^+\pi^-$, 
reactions in which the final $e^+$ and $e^-$ are {\it both} 
lost down the beam pipes
and, thereby, provide the large missing energy and missing mass
associated with the $\cnone\cnone$ pair in the signal reaction. If we
trigger on (i.e. tag) a photon with substantial $p_T^\gamma$,  either
the $e^-$ or the $e^+$ will be given a sufficient transverse
kick that it will be detected. To be precise,
we define $\theta_d$ to be the angle above which
an electron or positron can be detected. The 
largest transverse momentum, $p_T^{\rm max}$, that can be carried by the
final $\epem$ pair without one being detected
arises when the final $e^+$ and $e^-$ are coplanar,
have relative $\phi=0$ and have
polar angles with respect to their respective beams,
$\theta^+$ and $\theta^-$, 
both equal to $\theta_d$: $p_T^{\rm max}=(E^++E^-)\sin\theta_d$.
Using the constraint 
$E^+=\sqrt s-E^--E^\gamma-E^X$, we have 
$p_T^{\rm max}=(\sqrt s-E^X-E^\gamma)\sin\theta_d$. 
Meanwhile, the smallest $p_T$ of the the $\gam+X$ system in the final
state is $p_T^{\rm min}=p_T^\gam-p_T^X$, where $p_T^X$ denotes
the magnitude of the $X$ system transverse momentum.
$p_T^{\rm min}<p_T^{\rm max}$ is required to avoid detection of both
the final $e^+$ and $e^-$.  This condition can be rewritten in the form
\beq
p_T^\gam<{(\sqrt s-E^X)\sin\theta_d+p_T^X\over 1+\sin\theta_d}\,.
\label{ptgamcond}
\eeq
For the scenario under consideration, $E^X$ and $p_T^X$ are of order $\dmchi$
times a modest boost factor, and do not exceed 1 GeV for the
scenarios of interest. Thus,
if we impose a $p_T^\gam$ cut that exceeds $\sqrt s {\sin\theta_d\over 1+\sin
\theta_d}$ then this background process cannot contribute. Current detector
plans are consistent with instrumenting the beam hole down to 
$\theta_d\sim 1^\circ$, for which a cut of $p_T^\gamma>0.0172 \sqrt s$
will suffice. For $\sqrt s=600\gev$, we would require $p_T^\gamma>10.3\gev$,
which motivates our nominal minimal cut of $p_T^\gamma>10\gev$.
With this cut, the only way in which an $\epem\to\gam+\epem+X$ process
can contribute to the $\gam+\pi^+\pi^-+\emiss$ final state
of interest is if $X$ itself
contains missing energy of magnitude approaching the
magnitude of the minimum $p_T^\gam$ being required. 
One candidate is
$X=\tau^+\tau^-\to \pi^+\pi^-\nu\anti\nu$, where the $\tau^+\tau^-$
invariant mass is large enough that the mass and momentum
of the $\nu\anti\nu$ system is at least 2 to 3 GeV.  However, the
$\pi^+$ and $\pi^-$ will tend to be much more energetic than
expected in the signal reaction (unless $\dmchi>500\mev$) and will tend
to be produced more forward and backward (with correlations
with the beam directions opposite that expected for the signal) because of the
$t$-channel $\tau$ exchange in the Feynman diagram that gives rise to 
$\gam^*\gam^*\to \tau^+\tau^-$. Further, the background from this
process can be independently measured using the $\tau\to\rho\nu$ decay modes,
and then subtracted. Another candidate is $X=\pi^+\pi^- Z(\to
\nu\anti\nu)$ for which $\epem\to\gam+\epem+X$ has a small cross section.
In events of this type, virtualities are large and 
the $\pi$'s produced will tend to be very energetic and largely
forward and backward, whereas the $\pi$'s from the signal reaction are
quite central and quite soft in the lab frame. 
Cuts that require soft central pions will thus
be very effective in eliminating such backgrounds while retaining
almost all the signal events. 

A second type of two-photon background arises from overlapping
$\epem\to \gam \emiss$ (e.g. $\gam Z^*(\to \nu\anti\nu)$,
with $m_{Z^*}>\mchichi$) and 
$\epem\to\epem\gam^*\gam^*\to\epem\pi^+\pi^-$ collisions
in which the final $\epem$ of the 2nd reaction are allowed to disappear down
the beam lines regardless of $p_T^\gam$.
At TESLA, with large bunch separation, the
two $e^-$'s must come from the same $e^-$ bunch and the
two $e^+$'s from the same $e^+$ bunch.
Note that a cut requiring large
$\emiss$, in particular large $\mslash\sim\mchichi$, 
can only be satisfied if the $\emiss$
comes from the same reaction as the $\gam$. The `hard' process will 
have a small cross section. Further, the $\pi$'s 
from the overlapping two-photon event tend
to be very energetic and the cut requiring central/soft $\pi$'s 
will be very effective (and essentially 100\% efficient for the signal events).
Further, the $\pi^+$ and $\pi^-$ produced in the 2nd collision will tend
to balance each other in transverse momentum, whereas this will
never be the case for the $\pi$'s from the signal events.
In addition, for most such events there would be substantial separation
in $z$ (displacement along the beam line) between the two overlapping
events. The LC vertex detectors will have exceptional ability
to determine the $z$ vertex location of the event producing the $\pi$'s,
assuming $c\tau$ for the $\cpmone$ decay is not large. (If it is large,
observation of large transverse impact parameters for
the $\pi$'s would, alone,
be sufficient to eliminate any background.) 
The $z$ location of the collision producing the $\gam$ 
will hinge on electromagnetic calorimetry and will
require a pre-shower plane for good accuracy. We are uncertain
of what resolution to expect here, but it could be sufficiently good that
a non-zero $z$ separation would be seen for most overlapping
collisions, which could be used to eliminate most of this background.

Another cut that is useful\footnote{We thank H.-U. Martyn
for pointing this out.}  for eliminating two
photon backgrounds is to require that the missing energy vector
$\vec{\pslash}=\vec p_{e^+}+\vec p_{e^-}-\vec p_{\gam}-\vec p_{\pi^+}
-\vec p_{\pi^-}$ have an angle of $> 100$ mrad with respect to
both beam axes.  In fact, our cuts of $p_T^\gam>10\gev$ 
and $10^\circ<\theta_\gam<170^\circ$ guarantee that
this is the case for all signal events; indeed, all but a tiny
fraction of the events have an angle between $\vec{\pslash}$ and
both beam axes of $>200$ mrad.

\begin{figure}[ht!]
\leavevmode
\begin{center}
\epsfxsize=6.25in
\hspace{0in}\epsffile{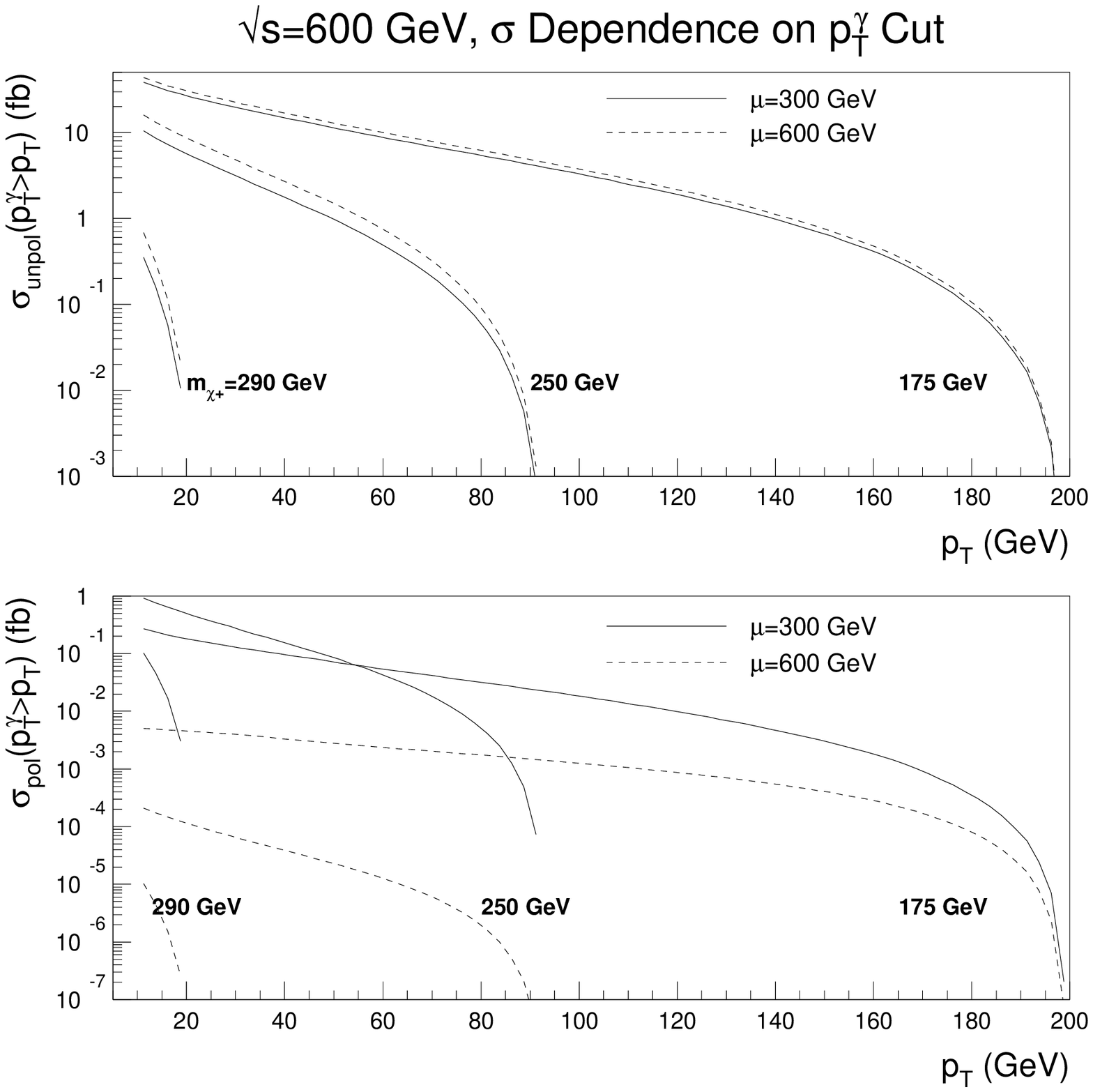}
  \caption{The $\rts=600\gev$
cross sections for $\gam\cpone\cmone$ production as a
function of the minimum photon $p_T^\gamma$. The
results for an unpolarized and a right-handed polarized $e^-$ beam
are shown in the upper and lower plots, respectively,
for the cases of $\mcpmone=175,250,290\gev$ and $\mu=300,600\gev$,
with $\tanb=5$. Cross sections are computed requiring
$10^\circ\leq\theta_\gamma\leq 170^\circ$.
These results assume that the sneutrino exchange diagram
can be neglected.}
\label{ptgam}
\end{center}
\end{figure}

In the absence of a reliable Monte Carlo for generating these
types of backgrounds, it is impossible to be absolutely certain
that a collection of small contributions will not add up to a
non-negligible result.  
The true background level will probably only be known once 
data taking has begun. If a background is found
for our minimal tag requirements 
of $p_T^\gamma>10\gev$, $10^\circ\leq\theta_\gamma\leq 170^\circ$
balanced by large missing mass and soft central pions, 
the $p_T^\gamma$ cut, can, for $\mcpmone$ not too near $\rts/2$,
be increased to the point where the background disappears.
For $L=1\abi$ of integrated luminosity, this will typically have little
impact on our ability to discover $\gam\cpone\cmone$ production
and determine $\mcpmone$ and $\dmchi$. In fact, one very good
set of procedures for the latter
determinations, discussed in a later section, focuses 
on events for which the photon energy is near maximal
and the invariant mass of the $\cpone\cmone$ system is very close
to $2\mcpmone$. In this kinematic region, backgrounds will surely
be negligible unless $\mcpmone$ is close to $\rts/2$ (implying that there
is no kinematic room for increasing the photon tag requirements).
To be more quantitative, we show in Fig.~\ref{ptgam} the unpolarized
beam and right-handed polarized $e^-$ beam cross sections
as a function of the minimum $p_T^\gamma$ accepted. Results
are shown for the representative
cases of $\mcpmone=175,250,290\gev$ and $\mu=300,600\gev$. In the unpolarized
case, $L=1\abi$ gives acceptable event rates for discovery
(we assume 10 events will be adequate if there is no background)
even for very large $p_T^\gamma$ cuts so long as $\mcpmone\lsim 275\gev$.
As discussed in a later section, these
event rates for large $p_T^\gamma$ 
are also quite adequate and, indeed, very useful for precise determinations
of $\mcpmone$ and $\dmchi$ for such $\cpmone$ masses.
(The ability to determine $\mcpmone$ from the event rate as a
function of the $p_T^\gamma$ cut is immediately apparent from the figure.)
At $\mcpmone=290\gev$, a cut of $p_T^\gamma >17\gev$ would
still leave a (discoverable rate) 
of 30 events (for $L=1\abi$) if the background
is negligible. Determinations of $\mcpmone$ and $\dmchi$
would become less accurate at such a high mass.
In the polarized case, event rates are very
sensitive to the $\mu$ parameter.  This is good in that
it provides important sensitivity to the $\mu$ (and $M_2$) SUSY parameters,
as described in a later section. However, Fig.~\ref{ptgam} makes
clear that the polarized cross sections are not necessarily large
enough to be accurately measured and might not be observable at all.
Use of the polarized cross sections will, in many cases,
require that the background
really is very small starting at $p_T^\gamma\sim 10-20\gev$.
Finally, we note that the results of Fig.~\ref{ptgam} assume
that the amplitude contribution from
the sneutrino exchange diagram is small. If the sneutrino mass
is at or below $\rts$, there will be substantial suppression
of the cross section due to the $\snu_e$ exchange diagram.

\begin{figure}[ht!]
\leavevmode
\begin{center}
\epsfxsize=4.25in
\hspace{0in}\epsffile{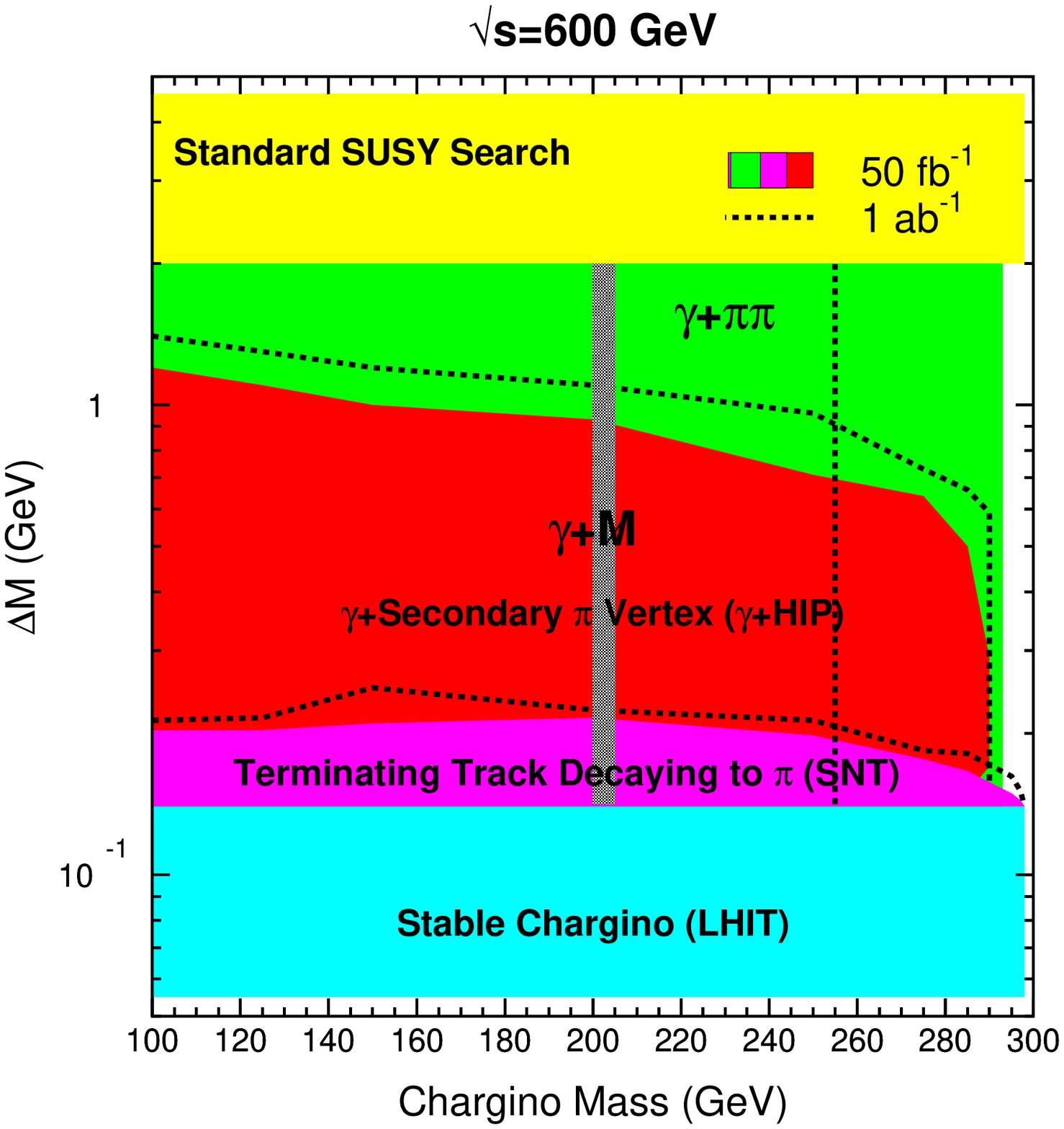}
  \caption{Regions in $\dmchi$, $\mcpmone$
parameter space for which different discovery modes are viable.
The SNT, $\gam+$HIP and $\gam+\pi\pi$ modes are presumed to be 
background free; we assume that 10 events will be adequate 
for discovery.  For the SNT,
$\gam+$HIP and $\gam+\mslash$ signals, discovery reach is given
for $L=50\fbi$ and also for $L=1\abi$. Also shown by the  vertical
band and vertical line (the band for $L=50\fbi$ and the line for $L=1\abi$)
is the reach of the $\gamma+\mslash$ detection mode, which
is relevant only if the soft $\pi$'s are not detectable, as might
be the case for some range of $\dmchi\gsim 200\mev$. (Limits
in this mode assume $S/\sqrt B>5$ {\it and}
$S/B>0.02$ is required; see text.) Note the large increase
in discovery reach for the $\gam+\mslash$ mode with increased $L$.}
\label{regions}
\end{center}
\end{figure}

We reemphasize the fact, apparent explicitly in
Fig.~\ref{regions} (discussed in the following section),  that there is a high
probability that one or both of the
pions in the $\gam\cpmone\cmone$ final state will have an observable
impact parameter whenever $\dmchi$ is below $\sim 1\gev$ and
$\mcpmone$ is in the $\mcpmone\lsim 200\gev$ region, as preferred
for the ($\delgs=0$ O-II)/AMSB boundary conditions.  In this case,
backgrounds will unquestionably be negligible and all the parameter
determination studies described later will be possible with the
described precisions.

Finally, we return to the possibility of using $\epem\to \wmp\cpmone\cnone$
in which the $W$ can decay either leptonically or hadronically and
$\cpmone\to \pi^\pm\cnone$. For moderate $\mcpmone$, the cross
section for the production process is quite substantial.
For example, for $\mcpmone=175\gev$, $\tanb=5$, $\mu=600\gev$
and $\rts=600\gev$ ($450\gev$),
the unpolarized beam cross section is 12 fb (0.46 fb) and
the pure $e_R^-$ cross section is 0.28 fb (0.007 fb),
after summing over both $\wm\cpone\cnone$ and $\wp\cmone\cnone$
production. 
The $\rts=600\gev$ cross sections (for instance) are to be
compared to the $p_T^\gam>10\gev$ cross sections of Fig.~\ref{ptgam}:
38.5 fb for unpolarized beams and 0.27 fb for pure $e_R^-$. 
The $\wmp\cpmone\cnone$ cross section is thus
substantial so long as $2\mcpmone+\mw$ is not too near $\rts$. This 
production mode could certainly be considered as an alternative 
 for discovery and parameter studies if the $\gam\cpone\cmone$
mode should (contrary to our expectation) have
significant background. We have not studied possible
backgrounds to the $\wmp\cpmone\cnone$
channels, although we expect backgrounds to be small.
We will only remark on its possible utility in a few places.

\noindent{\bf 3. Discovery Reach.}
Following the above discussion, we can 
estimate the regions in $(\dmchi,\mcpmone)$
parameter space for which the various SUSY discovery 
modes will be viable at a $\rts=600\gev$ LC. They 
are shown in Fig.~\ref{regions}.
For $\dmchi\gsim 2\gev$, we assume (following the LEP experience)
that observation of $\cpone\cmone$ production will be possible
up to very nearly $\mcpmone\sim \rts/2$. For $\dmchi<m_\pi$,
the $\cpmone$ will be sufficiently stable in the detector
that an LHIT signal for $\cpone\cmone$ production will be easily detected up
to threshold. For $\dmchi$ between $m_\pi$ and roughly $200\mev$, the SNT
signal will be viable.  In fact,
since triggering will not be necessary
at the NLC, even (STUB) charged tracks as short as 4 or 5 cm can be directly 
imaged,~\footnote{We thank M. Peskin for bringing this to our
  attention.} 
implying that our $\beta<0.8$ heavy ionization requirement for the SNT
signal could be relaxed. For $200\mev<\dmchi<2\gev$, it is very probable
that one will need to employ the $\gam$ tag signatures, i.e. 
$\gam+$HIP(s) and $\gam+\pi$(s) (always with large $\emiss$).
We believe that the $\gam+$HIP and SNT signatures will be background-free.
(They disappear at the largest values
of $\mcpmone$ due to inadequate boost for the produced $\cpmone$,
which leads to too small an impact parameter for the HIP or too short
a path length for the STUB track.) 
Development of Monte Carlo programs that can accurately compute
the backgrounds to the above signals should be a high priority,
especially for the $\gam+\emiss+\pi$(s) ($\gam+\pi$(s) for short) signal.
However, the measured background after cuts at LEP2 for the $\gam+\pi$(s) 
signal was negligible or very small
even without requiring a high impact parameter for at least
one of the $\pi$'s. This, coupled with the arguments given
in the previous section allow for some optimism
that the background for the $\gam+\pi$(s) 
signal will continue to be small at high $\rts$, even for the relatively mild
$p_T^\gamma>10\gev$ cut. 
The exact boundaries for the LHIT, SNT, $\gam$+HIP(s) and $\gam+\pi$(s) 
signatures shown assume that 10 events 
are adequate to detect the above signals (assuming no background).
The rapid $s$-wave threshold turn-on of
$\gam\cpone\cmone$ production allows one to probe
almost to the $\mcpmone\sim \rts/2-p_T^\gamma$ kinematic limit
so long as the detector can resolve
the soft $\pi$'s from the $\cpmone$ decays.
As shown in Fig.~\ref{regions}, 
even assuming zero background, $L=1\abi$ does not greatly
increase the discovery reach of the STUB and HIP channels as compared
to $L=50\fbi$.
The increase in reach for $L=1\abi$ for the (stable) LHIT signal is very small 
and is not shown. Since it is possible that backgrounds that are negligible
at $L=50\fbi$ result in a non-negligible
number of events at $L=1\abi$, one should not absolutely rely
on the indicated increase in coverage for the SNT and $\gam+$HIP signals
for the latter $L$.

\begin{figure}[ht!]
\leavevmode
\begin{center}
\epsfxsize=6.25in
\hspace{0in}\epsffile{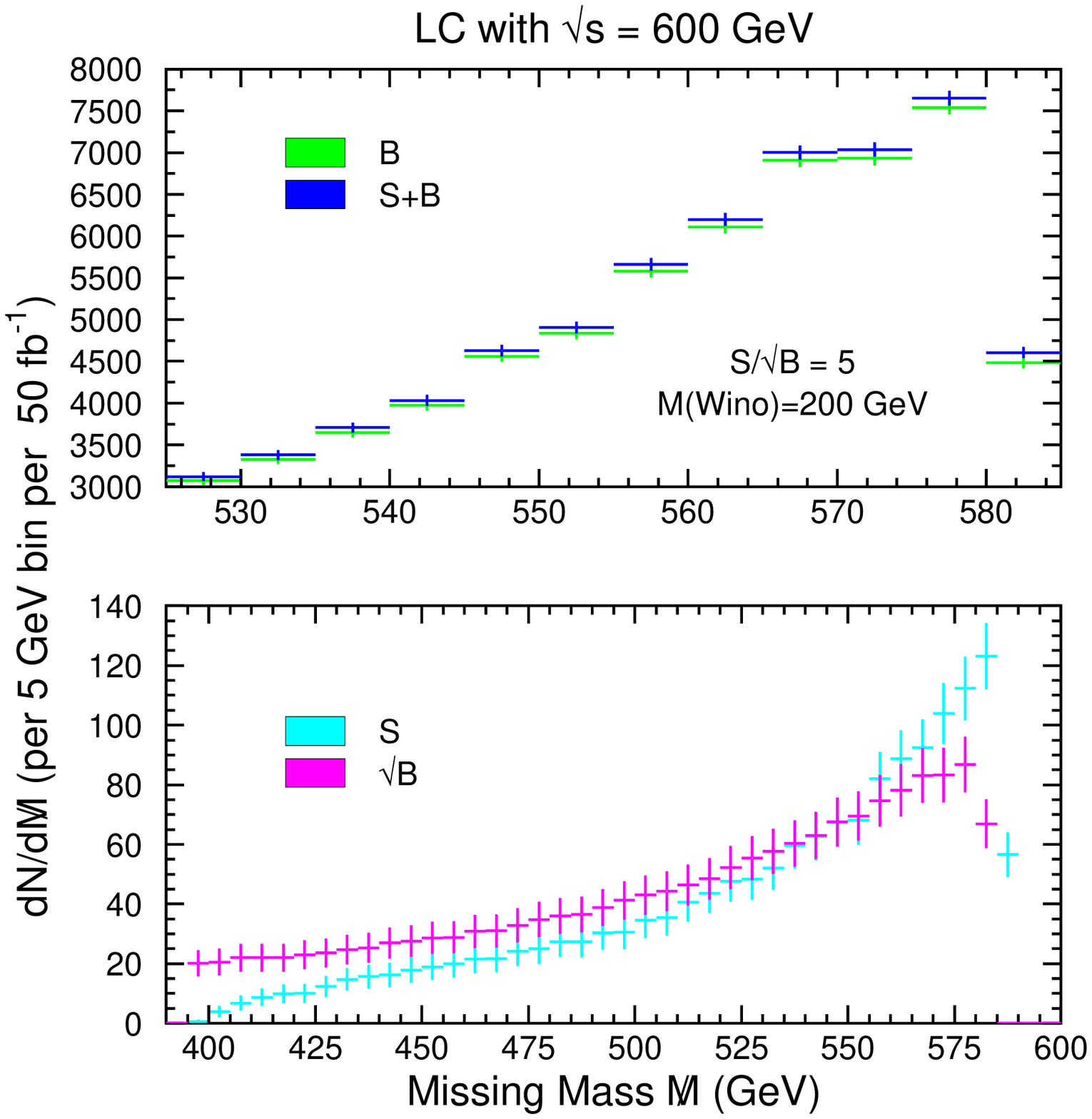}
  \caption{$dN/d\mslash$ for signal and background in
the $\gam+\mslash$ final state using 5 GeV bins
and assuming $L=50\fbi$.  Also shown are $S$ and $\sqrt B$.}
\label{dndmslash}
\end{center}
\end{figure}

Given that the boundary conditions in models 
with loop-dominated gaugino masses
are such that $\dmchi\in[200\mev,2\gev]$ is almost a certainty, 
the reach of the
$\gam$-tag+$\emiss$+soft-$\pi$'s (with or without HIP(s))
signal is of great importance.   
If the soft pions cannot be detected/tracked before they curl up (e.g.
because the magnetic field is too strong), then sensitivity
to $\gam\cpone\cmone$ production will be much more limited.
(This is illustrated by the somewhat higher
$\mcpmone$ limits of L3 vs. DELPHI for  $\dmchi\in[200\mev,2\gev]$
due to the fact that L3's magnetic field is weaker than DELPHI's.)
If the $\pi^\pm$ tracks are not detected, one important signal is 
$\epem\to \gamma\cpone\cmone\to\gamma+\mbox{invisible}$, for which the
$\epem\to\gamma \nu\anti \nu$ background is
substantial. For any given presumed value of $\mcpmone$,
we place a cut on the invisible mass, $\mslash$, recoiling opposite
the trigger photon of at least $\mslash>2\mcpmone$, 
that removes much of the $\nu\anti\nu$ background, 
in particular that from on-shell $\gamma+Z$ production.
The more limited discovery reach is illustrated
by the $\gam+\mslash$ boundaries (one for $L=50\fbi$
and one for $L=1\abi$) drawn at fixed $\mcpmone$ for 
$\dmchi\in[200\mev,2\gev]$
in Fig.~\ref{regions}. For lower $\mcpmone$ values, cuts
can be placed on $\mslash$ so as to obtain 
$S/B\geq 0.02$ and $S/\sqrt B\geq 5$ for $L= 50\fbi$. 
The difficulty  of observing the signal for large $\mcpmone$ can be seen from 
the $\mcpmone=200\gev$, $L=50\fbi$ 
results shown in Fig.~\ref{dndmslash}. The upper part of the figure
gives $S$ and $S+B$ in 5 GeV bins in the $\mslash>525\gev$ region. The lower
part of the figure shows $S$ and $\sqrt B$ 
for 5 GeV bins. We see that $S/\sqrt B$ per bin is only of
order 1 for $\mslash> 525\gev$. Although at the peak in $S$,
located at $\mslash\sim 580\gev$, the 5 GeV bin
has $S/B=0.028$, for $L=50\fbi$ this one bin only yields 
$S/\sqrt B=1.87$. Thus, for 
this $L$ it is necessary to include 
all the bins with $\mslash\geq 525\gev$
in order the get a net $S/\sqrt B>5$ signal; in this 
case the net $S/B$ is $\sim 0.02$. However, with higher $L$ one
can do much better by focusing on the highest $\mslash$
bins for which $S/B$ is largest.
\begin{table}[h!]
\renewcommand{\arraystretch}{.7}
  \begin{center}
    \begin{tabular}[c]{|c|c|c|c|c|c|c|c|} \hline
$\mcpmone$ (GeV) & 175 & 200 & 250 & 260 & 275 & 285 & 290 \\
\hline
$\sigma$ (fb) & 3.9 & 3.6 & 2.3 & 1.8 & 1.4 &  0.90 & 0.52 \\
$S/B$ & 0.043 & 0.040 & 0.025 & 0.020 & 0.016 & 0.010 & 0.006 \\
$L$ (fb$^{-1}$) for $5\sigma$ & 150 & 175 & 435 &  690 & 1140 & $\sim 3000$ & $\sim 8500$ \\
\hline
    \end{tabular}
    \caption{$\gam+\mslash$ results for $580\leq \mslash\leq 590\gev$.}
    \label{smslash}
  \end{center}
\end{table}
In Table~\ref{smslash} we give the $\gam+\mslash$ cross section
$\sigma$ as a function
of $\mcpmone$ for $580\gev\leq\mslash\leq 590\gev$, the
ratio $S/B$,  and the luminosity required for $S/\sqrt B=5$.
From this table, we see that if $S/B\gsim 0.02$ is required for 
a viable signal, then $\mcpmone$ up to $\sim 260\gev$ can be
probed for $L\geq 700\fbi$.  If systematics can be
controlled to the $S/B\sim 0.01$ level, then the $\gam+\mslash$
signal could be observed 
up to $\mcpmone\sim 285\gev$ for very high $L\sim 3000\fbi$.

\noindent{\bf \boldmath 4. Determining $\mcpmone$ and $\dmchi$.} 
If the chargino or its decay products are directly visible,
for expected $\epem$ collider luminosities it will be possible
to detect $\cpone\cmone$ or $\gamma\cpone\cmone$ production for $\mcpmone$ up
to values very close to the $\rts/2$ threshold. 
The soft $\pi$(s) from the $\cpmone$ decay or the non-prompt
$\cpmone$ decay LHIT, SNT and/or $\gam+$HIP signals will indicate clearly
that $\dmchi$ is small. The next important task will
be to measure $\mcpmone$ and $\dmchi$ as precisely as possible.
We focus entirely on scenarios with $\mpi<\dmchi\lsim 0.8\gev$ for which
we must employ the $\gam\cpone\cmone$ channel and for which
the $\cpmone\to \pi^\pm\cnone$ branching ratio is close to 1. 
For larger $\dmchi$,
one would have to take into account the $\cpmone\to \pi^\pm\pi^0\cnone$
and $\cpmone\to\ell^\pm\nu\cnone$ decays.  Our techniques could still
be applied, but the effective event rates for the $\pi^\pm\cnone$
channels would be reduced and some accuracy lost unless the other modes
could be included in the analysis following analogous procedures
(which are relatively straightforward extensions of those presented). 

\begin{figure}[ht!]
\leavevmode
\begin{center}
\epsfxsize=6.25in
\hspace{0in}\epsffile{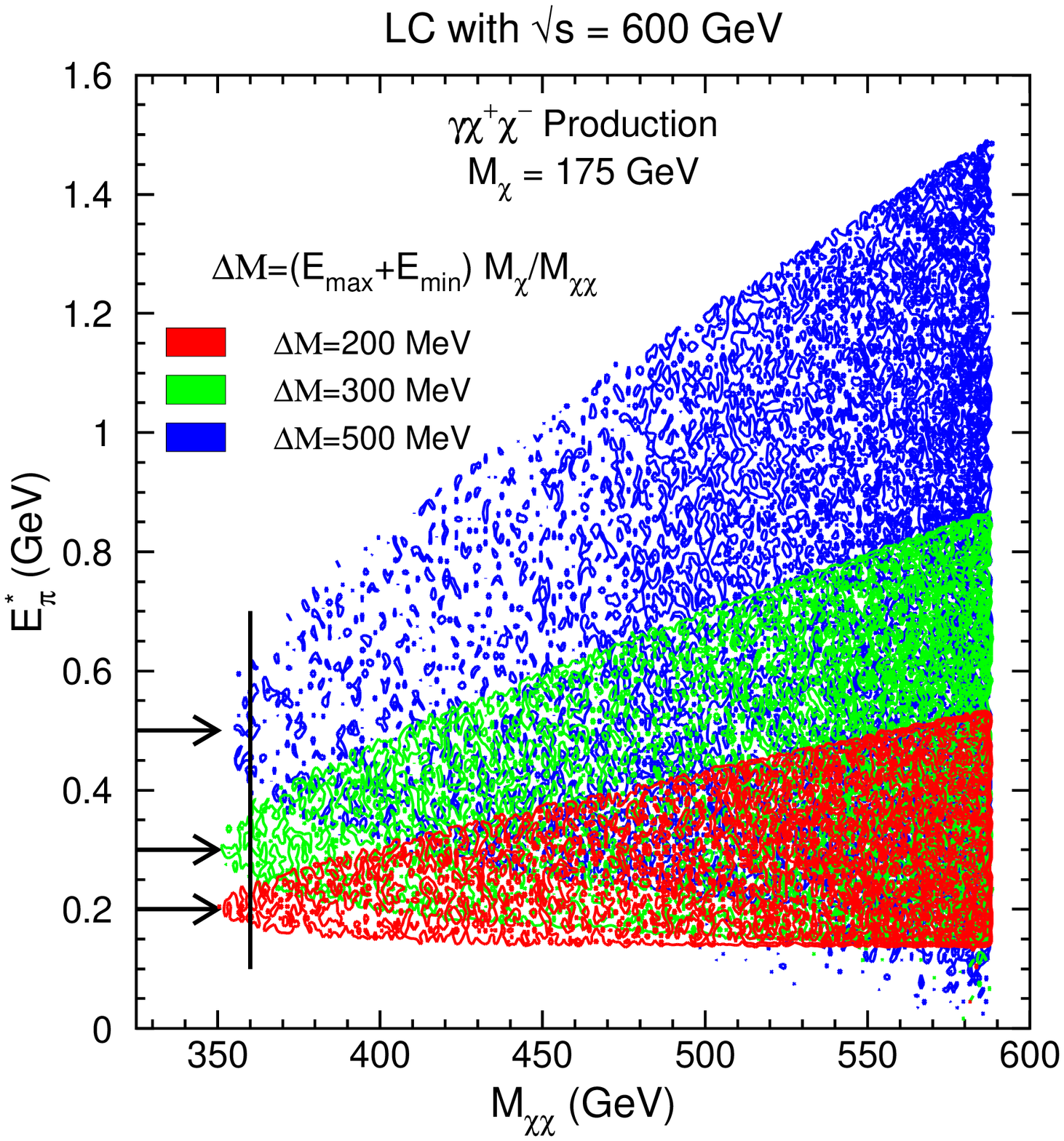}
  \caption{We plot event number contours in $[\mchichi,E_\pi^*]$
parameter space for several different values of $\dmchi$
taking $\mcpmone=150\gev$. Note the event concentration at high $\mchichi$.
The arrows indicate the values of $\dmchi$, which are approximately equal
to $\epipr$ at the $\mchichi=2\mcpmone$ endpoint. The vertical
line at $360\gev$ denotes the $\mchichi$ value below which the 
$\epipr$ distribution is approximately Gaussian.}
\label{pidist}
\end{center}
\end{figure}

The most direct technique for determining $\mcpmone$ and $\dmchi$
independently of theoretical assumptions is via
the $\pi^\pm$ kinematic distributions. We shall later discuss
the determination of $\dmchi$ via observing a distribution
of $\pi^\pm$ impact parameters. By way of preview, we note
that the impact parameter
distribution is determined primarily by $c\tau$ and $\dmchi$,
with weak dependence on $\mcpmone$.
Once $\mcpmone$ is determined via the kinematic procedures
we shall describe below, in favorable cases (large $c\tau$)
the impact parameter distribution
will allow a simultaneous determination of $c\tau$ and $\dmchi$
with errors on $\dmchi$ that are competitive with the errors
we shall find using kinematic distributions.
Of course, the kinematic distribution procedures we now describe
are essential when the $\cpmone$ decay is simply
too prompt to directly observe its path length.
However, even if the distribution of path lengths
is not very apparent (because most events are clustered at
low impact parameter, e.g. as typical for $\dmchi\sim 800\mev$ for which
$c\tau\sim 190~\mu$m) or useful, most events {\it will} have
an observable non-zero impact parameter for at least one $\pi$
and, consequently, backgrounds to the $\gam+\pi^+\pi^-+\emiss$ signal
will certainly be negligible after our simple cuts
on $p_T^\gam$, $\theta_\gam$ and $\mchichi$ described earlier. 

The observables in the $\gam\cpone\cmone\to\gam\pi^+\pi^-\cnone\cnone$
final state of interest are the invariant mass of the chargino pair, 
$\mchichi\equiv\sqrt{(p_{e^+}+p_{e^-}-p_\gamma)^2}$, and
the energies, $\epipr$, of the $\pi$'s in the $\mchichi$ rest frame.
For $\mchichi>2\mcpmone$, the $\cpmone$ are boosted in the $\mchichi$
rest frame and 
$\epipr$ depends upon the orientation of $\vec p_\pi$ in the $\cpmone$
rest frame with respect to the boost direction. 
The maximum and minimum $\epipr$ values are given by
\beq
E_\pi^{\star~{\rm max}}=\gamma_{\cpmone}(E_0+\beta_{\cpmone} p_0)\,,\quad
E_\pi^{\star~{\rm min}}=\gamma_{\cpmone}(E_0-\beta_{\cpmone} p_0)\,,
\eeq
with 
\beq
\beta_{\cpmone} =
\sqrt{1-{4\mcpmone^2\over\mchichi^2}}\,,\quad 
\gamma_{\cpmone}={1\over\sqrt{1-\beta_{\cpmone}^2}}\,,\quad
E_0={\mcpmone^2+\mpi^2-\mcnone^2\over 2\mcpmone}\simeq \dmchi\,,\quad
p_0=\sqrt{E_0^2-\mpi^2}
\label{boostsetc}
\eeq
characterizing the charginos in the $\mchichi$ rest frame
and the $\pi^\pm$ in the $\cpmone$ rest frames. 
These equations can be inverted to give
\bea
\gamma_{\cpmone}^2&=&{\ebar^2+\mpi^2-\ehat^2
\pm\sqrt{(\ebar^2+\mpi^2-\ehat^2)^2-4\ebar^2\mpi^2}\over 2\mpi^2}\,,
\label{inversiona}
\\
\mcpmone&=&{\mchichi\over 2\gamma_{\cpmone}}\,,\quad
\dmchi=\mcpmone-\sqrt{\mcpmone^2-{2\mcpmone\ebar\over\gamma_{\cpmone}}+\mpi^2}
\simeq {\ebar\over\mcpmone}\,,
\label{inversionb}
\eea
where $2\ebar=E_\pi^{\star~{\rm max}}+E_\pi^{\star~{\rm min}}$
and $2\ehat=E_\pi^{\star~{\rm max}}-E_\pi^{\star~{\rm min}}$.
(The $\simeq$ expressions in Eqs.~(\ref{boostsetc}) and (\ref{inversionb})
apply if we neglect terms of order $\mpi^2/\mcpmone^2$
and $\dmchi^2/\mcpmone^2$.)
For any given $\mchichi$, there is a 
two-fold ambiguity in $\gamma_{\cpmone}^2$. The $-$ ($+$) sign applies
for lower (higher) $\mchichi$ values. Only one set of choices as
a function of $\mchichi$ will lead to $\mchichi$-independent values for
$\mcpmone$ and $\dmchi$. Even at a fixed $\mchichi$, fitting the actual
distribution in $\epipr$ can provide a unique determination of $\mcpmone$
and $\dmchi$, given enough statistics.  

Since the $\pi$'s are soft,
the resolution for measuring $p_\pi$ in 
the lab frame will be dominated by the `constant' term, which has a typical 
value of $\sim 0.5\%$. Further,
$\delta\mchichi/\mchichi\sim   
\delta E_\gam\rts/\mchichi^2\sim 0.1\sqrt{E_\gam}
\rts/\mchichi^2$ (all in GeV)
will also be very small. We have generated events including
the measurement smearing in $E_\gam$. For each
event with a given $\mchichi$, we boost to the $\mchichi$ (i.e. chargino-pair) 
rest frame
and compute the energies, $\epipr$, of the observed $\pi$'s.
In Fig.~\ref{pidist}, we show the region of the $[\epipr,\mchichi]$
plane occupied by the events for several different choices of $\dmchi$,
taking $\mcpmone=175\gev$. 
For $L=50\fbi$ and $\rts=600\gev$,
the occupied region will contain $\sim 5000$ entries (2 entries per event). 
A large fraction of the entries reside near the large-$\mchichi$ boundary.

The location of the threshold in $\mchichi$ provides a particularly 
useful way to measure $2\mcpmone$. The statistical error 
for such a determination depends upon the
number of events near the threshold.
We assume that the $\gam+\pi\pi$ signal is background-free and
compute the approximate $1\sigma$ error $\delta \mcpmone$
from the criterion:
\beq
{\what S(\mcpmone-\delta\mcpmone)-\what S(\mcpmone) \over 
\sqrt{\what S(\mcpmone-\delta\mcpmone)}}=1\,,
\label{crit}
\eeq 
where $\what S$ is the number of events in the near-threshold
region $2\mcpmone\leq \mchichi\leq 2\mcpmone+10\gev$.
For $\mcpmone\in[150,225]\gev$ ($\in[250,275]\gev$) we find 
$\delta\mcpmone\sim 1\gev$ ($\sim 0.5\gev$) for $L=50\fbi$ and 
$\delta\mcpmone\sim 0.2\gev$ ($\sim 0.1\gev$) for $L=1\abi$.
The accuracy improves at higher masses because of the increasing
sharpness of $d\sigma/d\mchichi$ as a function of $\mchichi$.
\begin{figure}[h!]
\leavevmode
\begin{center}
\epsfxsize=6.25in
\hspace{0in}\epsffile{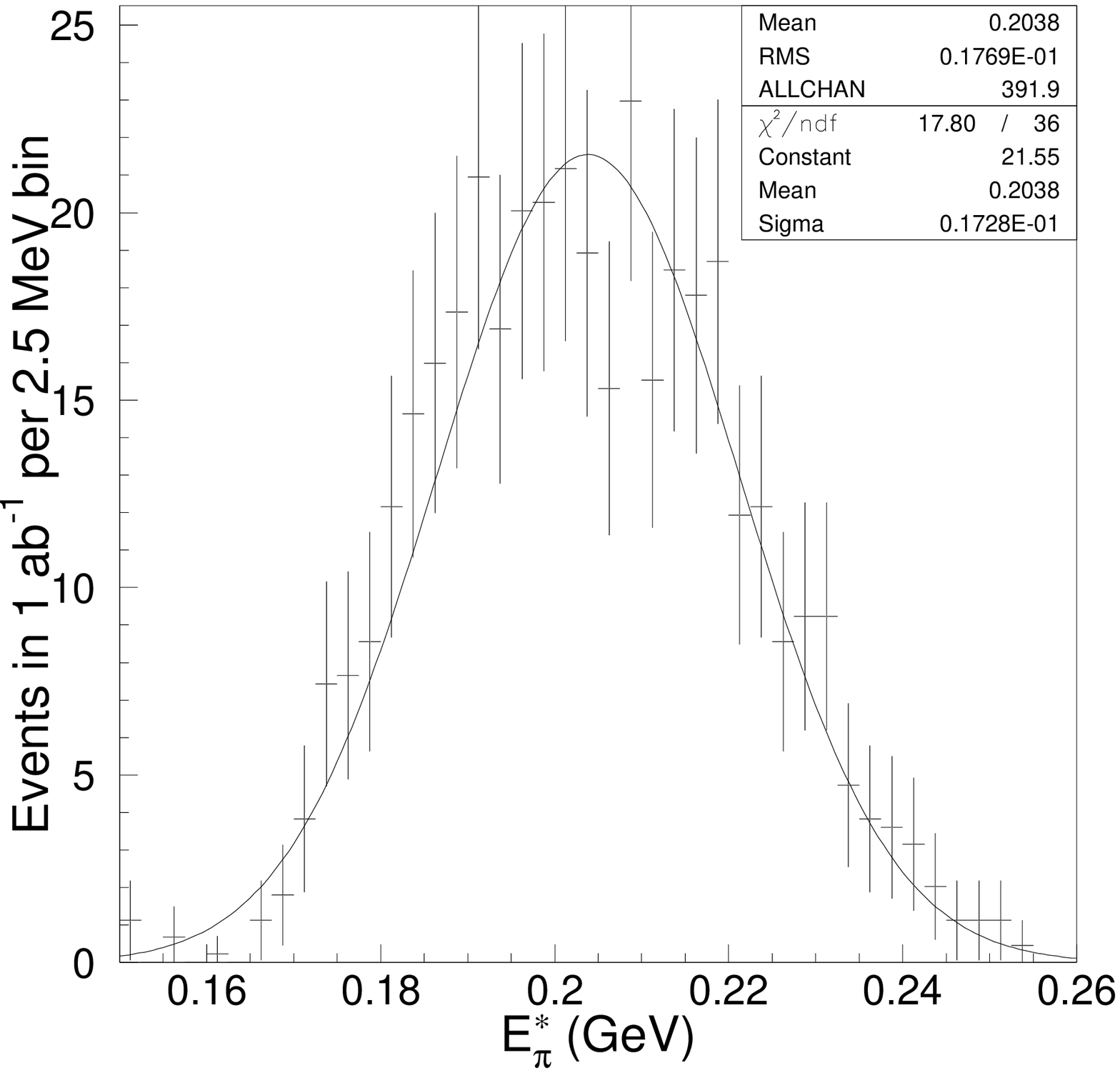}
  \caption{We plot the distribution of $\epipr$ for the
case of $\mcpmone=175\gev$, $\dmchi=200\mev$ and $\mchichi\leq 360\gev$.
Also shown is the Gaussian fit and its r.m.s. used to determine
the error for this case in Table~\ref{dmchierrors}.
We took $L=1\abi$, yielding 392 events in the plot.}
\label{epion}
\end{center}
\end{figure}

\begin{table}[h!]
\renewcommand{\arraystretch}{1}
  \begin{center}
    \begin{tabular}[c]{|c|c|c|c|c|} \hline
\ & \multicolumn{2}{c|}{$\mchichi$ cut I} & \multicolumn{2}{c|}{$\mchichi$ cut
II} \\
\hline
[$\mcpmone$ (GeV), $\dmchi$ (MeV)] & $\anti{\epipr}$ (MeV) & $\delta\anti{\epipr}$ (MeV)  & $\anti{\epipr}$ (MeV) & $\delta\anti{\epipr}$ (MeV) \\
\hline
\renewcommand{\arraystretch}{.5}
$[175,200]$ & 201.6 & 1.9 & 203.8 & 0.9 \\
$[175,300]$ & 303.4 & 3.4 & 306.7 & 1.8 \\
$[175,500]$ & 493.2 & 9.0 & 499.0 & 3.6 \\
\hline
$[225,200]$ & 200.6 & 1.3 & 203.2 & 0.7 \\
$[225,300]$ & 300.4 & 2.6 & 303.6 & 1.2 \\
$[225,500]$ & 493.9 & 6.8 & 498.3 & 2.7 \\
\hline
$[250,200]$ & 200.4 & 1.0 & 202.8 & 0.5 \\
$[250,300]$ & 299.9 & 2.1 & 303.5 & 1.0 \\
$[250,500]$ & 494.0 & 5.0 & 498.2 & 2.1 \\
\hline
    \end{tabular}
    \caption{Summary of $\anti\epipr$ values and errors, $\delta\anti{\epipr}$,
 as a function of $\mcpmone$,
$\dmchi$ and $\mchichi$ cut, for $\mcpmone=175$, $225$ and $250\gev$
and $\dmchi=200$, $300$, and $500\mev$, taking $L=1\abi$.}
    \label{dmchierrors}
  \end{center}
\end{table}

The most direct way to determine $\dmchi$ is to concentrate on the 
$\mchichi\sim 2\mcpmone$ region for which $\epipr\simeq E_0\simeq \dmchi$.
Because of resolution smearing, there are actually events with 
$\mchichi<2\mcpmone$ and a finite spread to $\epipr$ in this region.
We have examined the $\epipr$ distribution for two cuts: 
(I) $\mchichi<2\mcpmone$ and (II) $\mchichi<2\mcpmone+10\gev$.  For cuts (I), 
there are fewer events 
but, for $\dmchi\lsim 400\mev$, 
the $\epipr$ distribution is more closely centered on $\dmchi$.
A typical case is illustrated in Fig.~\ref{epion}, based on cuts (II).
The error on $\anti{\epipr}$, the average $\pi$ energy,  is 
given by $\delta \overline{\epipr}/\sqrt{N}$,
where $N$ is the number of events and $\delta \overline{\epipr}$ is estimated
as the width of a Gaussian fit to the $\epipr$ spectrum.
Table~\ref{dmchierrors} gives
$\anti{\epipr}$ and $\delta\anti{\epipr}$ for 
both $\mchichi$ cuts (I) and (II) for a number of $\mcpmone$ and $\dmchi$
choices assuming $L=1\abi$. $L=50\fbi$ is not adequate
to give particularly small errors, but would allow a first determination
of $\dmchi$ to within 10 to 20 MeV or so. 
There is considerable variation of $\delta\anti{\epipr}$
with the case and the cut.
In addition,
$\anti{\epipr}$ is usually not precisely equal to $\dmchi$,
with shifts ranging up to 7 MeV, depending on the exact case and cuts
employed. However, the expected shape 
for the $\epipr$ distribution for any given choices
of $\mcpmone$, $\dmchi$ and $\mchichi$ cut is precisely known 
(to the extent that
the resolutions for $E_\pi$ and $E_\gam$ are known) and the expected
shift $\anti{\epipr}-\dmchi$ can be computed. 
Our estimate is that one could in the end achieve an uncertainty
for $\dmchi$ of order 2--5 MeV or better in most cases.
Once actual data were available, one would take
the experimental distribution resulting from the underlying $\mcpmone$
and $\dmchi$ and compare it to a selection of theoretical predictions
for the measured $\mcpmone$ (using the technique described earlier) and
a range of $\dmchi$ choices and minimize the $\chi^2$.

To repeat a point from the introduction, we note that the
above $\mchichi\sim 2\mcpmone$ techniques for determining
$\mcpmone$ and $\dmchi$ (and, of course, detecting the charginos
in the first place) imply a focus on events with the most energetic
photons.  Unless $\mcpmone$ is quite near $\rts/2$, the typical
photon-tag energy and transverse momentum
will be large enough that two-photon (and certainly other)
backgrounds to our $\gam\pi^+\pi^-\emiss$ final state will
surely be negligible even if $\dmchi$ is large
enough that most events do not have an observable
HIP for one of the $\pi$'s. In contrast, 
the techniques discussed in the following
rely on using the full range of photon tag energies and transverse
momenta, and could be compromised if (contrary to our expectation)
there is significant background at lower $E_\gamma,p_T^\gamma$ values
when $\dmchi$ is such that HIP(s) are not observable.

\begin{figure}[ht!]
\leavevmode
\begin{center}
\epsfxsize=6.25in
\hspace{0in}\epsffile{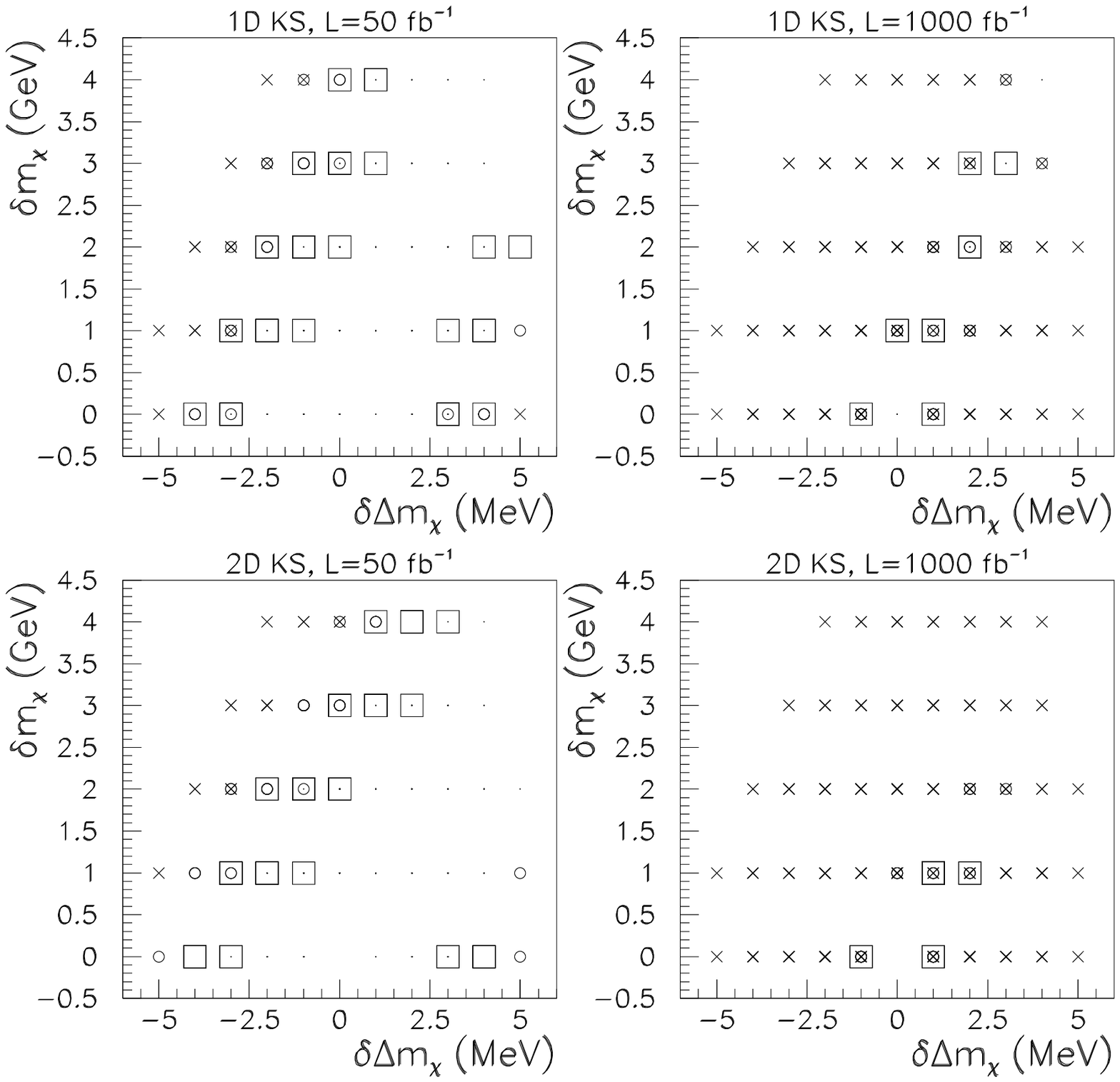}
  \caption{We show different levels of $\qks^{\rm 1D}$ 
and $\qks^{\rm 2D}$ computed from the $\epipr$ and $[\mchichi,\epipr]$
distributions (respectively) for two different choices, $A$ and $B$, 
of $\mcpmone$ and $\dmchi$,
with $173<\mcpmone<177\gev$ and $198<\dmchi<203\mev$,  as a function of
$\delta\mcpmone\equiv \mcpmone(A)-\mcpmone(B)$ in GeV (taken to be $>0$
by convention) and $\delta\dmchi\equiv \dmchi(A)-\dmchi(B)$ in MeV.
Results shown are for $L=50\fbi$ and $L=1\abi$. The symbols $\times$,
$\circ$, and $\square$ correspond to $\qks<0.1$, $0.1<\qks<0.3$,
and $0.3<\qks<0.68$, respectively. Overlap of symbols means that
different results are obtained for different choices of $\mcpmone(A)$
within the above range. Tiny dots are points that cannot be distinguished,
while the blank locations were not sampled in our analysis.}
\label{errors}
\end{center}
\end{figure}

We now compare the above $\mchichi\sim 2\mcpmone$
results to those obtained by employing
the full distribution in $[\mchichi,\epipr]$, including not only the
location of the end-points in the $\epipr$ spectra as a function
of $\mchichi$, but also the full shape of the distribution.
In particular, one could hope to make use of the large number of events
at large $\mchichi$. We now describe two procedures
for estimating the resulting errors on $\mcpmone$ and
$\dmchi$. In the first, we 
consider only the one-dimensional distribution in $\epipr$ as
a function of $\mcpmone$ and $\dmchi$.
We performed our study for $\mcpmone$ values
centered on $175\gev$ and $\dmchi$ values centered on $200\mev$
and employed bins of size $10\mev$ (i.e. chosen to
be somewhat larger than the resolution for $\epipr$). 
For two choices ($A$ and $B$) of $[\mcpmone,\dmchi]$, 
we found the maximum difference $\dmax$, based on the 1-D
Kolmogorov-Smirnov (KS) \cite{ks} procedure,
between the two cumulative distribution functions as
a function of $\epipr$ and computed the probability,
$\qks^{\rm 1D}\equiv P(D>\dmax)$, 
that the observed (or more precisely
for theoretically computed event rates, the expected) value of $\dmax$  
is inconsistent with the two distributions
having come from the same parent distribution rather than from
two different distributions. The values of $\qks^{\rm 1D}$ as a function
of $\delta\mcpmone\equiv \mcpmone(A)-\mcpmone(B)$ and $\delta\dmchi\equiv
\dmchi(A)-\dmchi(B)$ are given in Fig.~\ref{errors} for choices 
of $\mcpmone$ and $\dmchi$ in the ranges
$173<\mcpmone<177\gev$ and $198<\dmchi<203\mev$.
Due to the correlation apparent from Eq.~(\ref{inversionb}), which shows 
that a change in $\dmchi$ can be compensated by a corresponding
change in $\mcpmone$ without greatly affecting the $\epipr$ distribution,
we find that the more direct determination of $\mcpmone$ from the $\mchichi$
threshold region is crucial, especially for $L=50\fbi$. For $L=50\fbi$
($L=1\abi$), a determination of $\mcpmone$ to within 1 GeV (0.2 GeV)
will allow a determination of $\dmchi$ to within $\pm 5\mev$ ($\pm 1\mev$)
at the $\qks^{\rm 1D}=0.3$ (roughly 1$\sigma$ exclusion) level.

A generalized KS test can also be applied to the full two-dimensional
distribution in the $[\mchichi,\epipr]$ plane. (As above, we employ bin
sizes of $10\mev$ for $\epipr$. For $\mchichi$,  we employ
bins of size $5\gev$, again somewhat larger than expected resolution
error.) 
Results for the resulting $\qks^{\rm 2D}$ 
values as a function of $\delta\mcpmone$ and $\delta\dmchi$
are given in the bottom two windows of Fig.~\ref{errors}.
For $L=50\fbi$, we see that the statistics are somewhat marginal
for use of the 2D technique, but a $\pm 5\mev$ determination of
$\dmchi$ is possible if $\mcpmone$ is known to within 1 GeV. 
For $L=1\abi$ the 2D
KS test is superior to both the 1D KS test based on binning only in $\epipr$
and the near-threshold technique.
Using only the 2D KS test, we find that both $\mcpmone$ and $\dmchi$
can be measured with good precision. In particular,
the $\qks^{\rm 2D}>0.3$ region is confined to $\delta\mcpmone<1\gev$,
$-1<\delta\dmchi<2\mev$.  If we combine
the 2D KS analysis with the $\mchichi$ threshold determination
of $\mcpmone$, we do even better. In particular, the $\mchichi$ threshold
determination of $\mcpmone$ to within $0.2\gev$ in the $L=1\abi$
case, implies a reduced range of 
$-1<\delta\dmchi<1\mev$ at the $\qks^{\rm 2D}=0.3$ level.


\begin{figure}[!ht]
\leavevmode
\begin{center}
\epsfxsize=5in
\hspace{0in}\epsffile{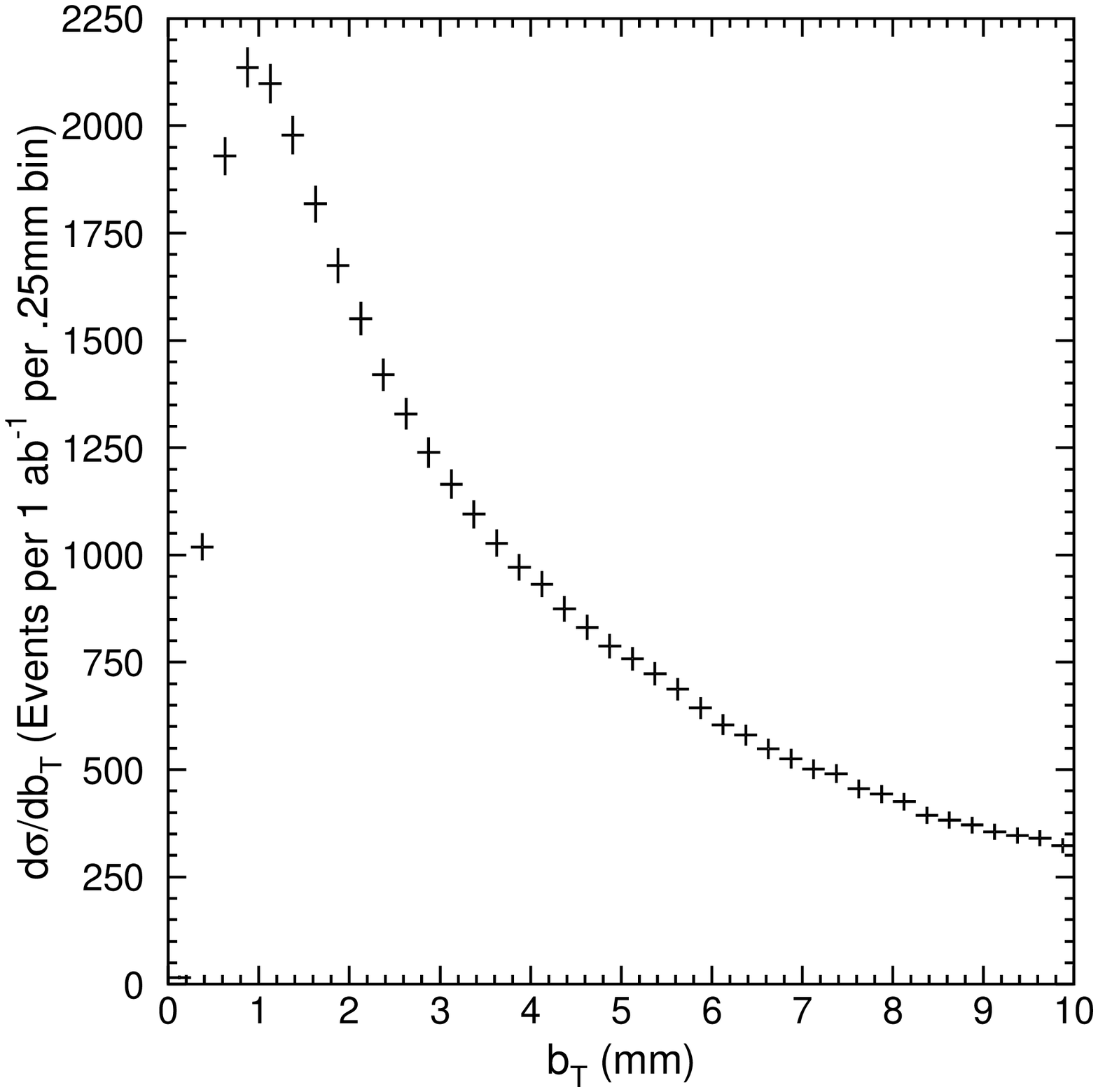} 
  \caption{For  $\rts=600\gev$, $\mcpmone=175\gev$,
$\dmchi=200\mev$ and $c\tau=23.81$ mm,
we plot the $b_T$ distribution for the final $\pi^\pm$
in $\epem\to\gam\cpone\cmone$ after the cuts described in the text.
The errors shown are those for $L=1\abi$ of integrated luminosity.
}
\label{btspectrum}
\end{center}
\end{figure}

Let us now return to the impact parameter distribution
for the $\pi^\pm$ in the final state
and the extent to which it can be used to 
determine $\dmchi$, $c\tau$ and underlying SUSY parameters.
For $\dmchi\lsim 800\mev$, the $\gam+$HIP(s)
signature typically has substantial rate.
We focus on the distribution of $\pi^\pm$ {\it transverse} 
impact parameters, $b_T$,
\footnote{Even though the exact location of the event vertex using
the $\gam$ trigger photon may be somewhat uncertain, $b_T$ can be reliably
measured as the distance in the transverse plane by which the projected
$\what p_\pi$ unit vector misses the $z$ axis.}
using the central reference
case of $\mcpmone=175\gev$, $\dmchi=200\mev$ and $c\tau= 23.81$ mm (the
nominal value from Table \ref{ctaus}, based on Ref.~\cite{cdg1}, 
for the given $\dmchi$ value).
We generate $\epem\to\gam\cpone\cmone\to\gam\pi^+\pi^-\emiss$
events at $\rts=600\gev$, using the minimal $p_T^\gam>10\gev$
cut, for given input values of $\mcpmone$, $\dmchi$
and $c\tau$. We decay the $\cpmone$ according to the chosen
$c\tau$, accounting for time-dilation for a given $\cpmone$
velocity on an event-by-event
basis. We compute the $b_T$ of the final state $\pi$'s if they
pass through the toy-model
vertex detector as specified in \cite{Gunion:2000jr},
which includes a first (L00-like) layer at 1.6 cm.
If $b_T$ is $\geq 5 \sigma_b$, where $\sigma_b$ is the
$p_T^\pi$-dependent impact parameter resolution given in 
\cite{Gunion:2000jr}, then we enter $b_T$ in the appropriate impact parameter
distribution bin. (There can be up to two entries per event.)
The resulting distribution for $\mcpmone=175\gev$,
$\dmchi=200\mev$ and $c\tau=23.81$~mm is plotted in
Fig.~\ref{btspectrum}. 
The bulk of the $b_T$ distribution lies in the $b_T<10$~mm range.
This is also the portion of the range for which fluctuations
will be under control for bins of reasonable size.  We have
chosen a bin size of 0.25 mm and, in what follows, we will consider only the bins
with $b_T<10$~mm.
Because of the complicated kinematics (including
the trigger photon), the limited size of the vertex detector,
and the $b_T\geq 5\sigma_b(p_T)$ requirement,
the shape of the $b_T$ distribution is not a simple
exponential, and, in particular, is cutoff at low $b_T$.
Assuming all these effects can be adequately modeled
and studied via Monte Carlo, the $b_T$ spectrum
can provide considerable information regarding the underlying parameters.
Indeed, the spectrum has a complicated dependence on $\mcpmone$
(as it affects the amount of $\pi$ boost),
$\dmchi$ (as it affects the amount of
momentum available for motion transverse to the $\cpmone$
direction) and $c\tau$.   Before the
$b_T\geq 5\sigma_b(p_T)$ requirement,
the  $b_T$ distribution shifts to larger $b_T$ with increasing 
$c\tau$, decreasing $\dmchi$ and decreasing $\mcpmone$.
However, the $b_T\geq 5\sigma_b(p_T)$ requirement removes more entries
at low $b_T$ for small $\dmchi$ than for large $\dmchi$, with the
result that larger $\dmchi$ actually has more weight at small $b_T$
than smaller $\dmchi$.  We now summarize the ability to use the $b_T$
distribution to determine $c\tau$ and $\dmchi$  for our detector
model. We reemphasize that use of this technique
requires that the acceptance and resolution of
the vertex detector and the influence of the precise cuts made all be
well understood.

\begin{figure}[!ht]
\leavevmode
\begin{center}
\epsfxsize=5in
\hspace{0in}\epsffile{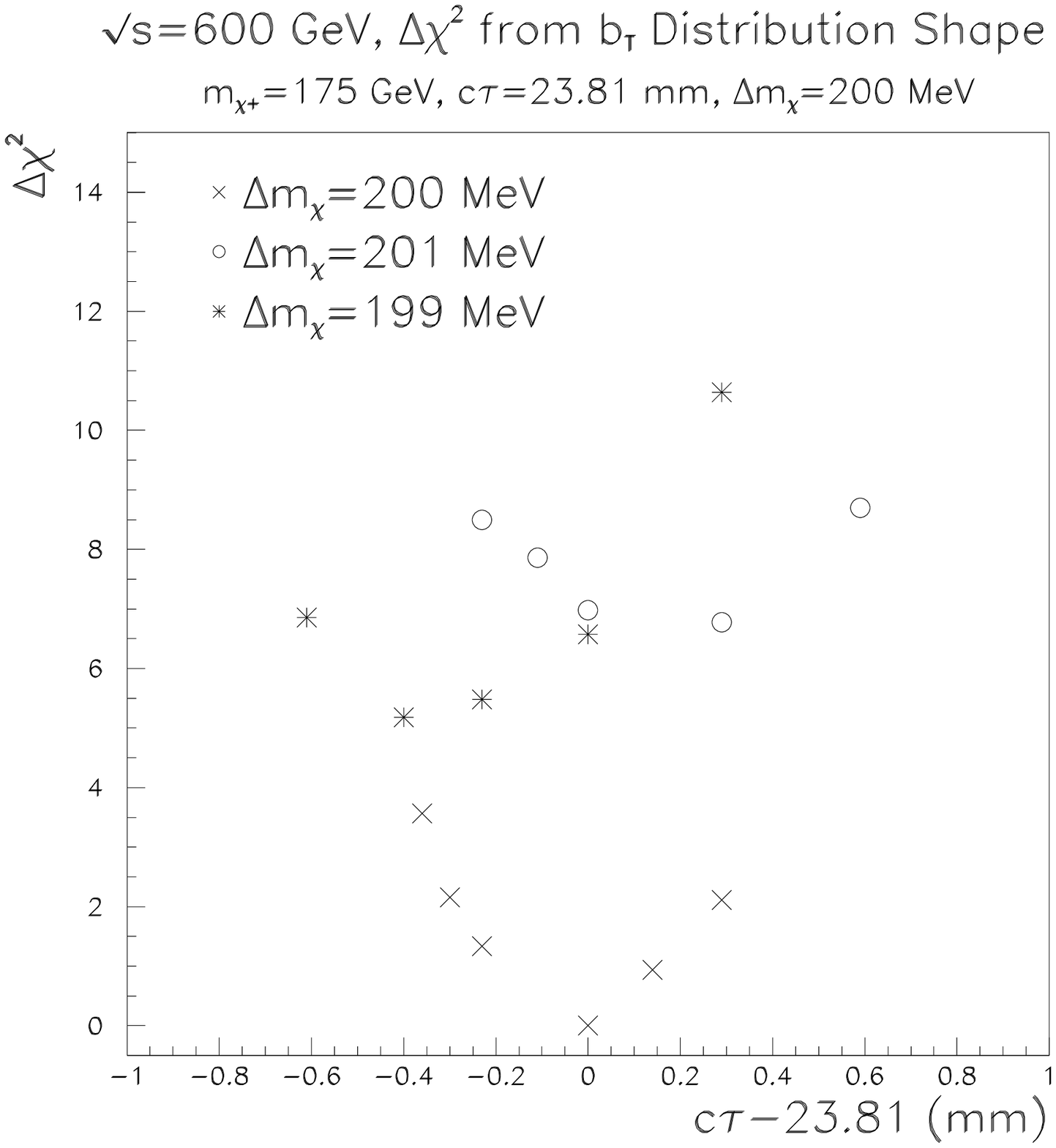} 
  \caption{For $\mcpmone=175\gev$, $\rts=600\gev$ and central values
of $\dmchi=200\mev$ and $c\tau=23.81$ mm,
we plot the $\Delta\chi^2$ computed as a function
of $c\tau-23.81~\mbox{mm}$ for $\dmchi=199$, $200$ and $201\mev$.
We have assumed $L=1\abi$, and have chosen the relative normalization
of the alternative distribution for each alternative parameter set
so as to minimize $\Delta\chi^2$.
}
\label{ctaudmchisq}
\end{center}
\end{figure}

To be quantitative, we have proceeded as follows.  We have chosen
the central values of $\dmchi=200\mev$, $c\tau=23.81$ mm 
and $\mcpmone=175\gev$. We then consider various neighboring values
and compute the $\Delta\chi^2$ between the $b_T$ distribution
for any given set of $[\dmchi,c\tau,\mcpmone]$ values
and the central values using bins of size 0.25 mm in the region
$b_T\leq 10$ mm and choosing the relative normalization that minimizes
$\Delta\chi^2$ (i.e. we rely on shape differences only).
The $b_T$ distribution is insensitive to changes in $\mcpmone$ within
the error of $\pm 0.2\gev$ from the kinematic distribution techniques.
Thus, it is most relevant to assess our ability to determine $c\tau$
and $\dmchi$ from the $b_T$ distribution.
Holding $\mcpmone=175\gev$ fixed, the $\Delta\chi^2$ values as a function
of $c\tau-23.81~\mbox{mm}$ are plotted for 
$\dmchi=199$, $200$ and $201\mev$
in Fig.~\ref{ctaudmchisq}.  For two parameters, the $1\sigma$ (68.27\%
CL) and $2\sigma$ (95.45\% CL) $\Delta\chi^2$ values are 2.30 and
4.61. We see that, assuming $L=1\abi$, 
a $\Delta\chi^2$ value below 4.61 can only be
achieved for $\dmchi$ values within $<1\mev$ from $\dmchi=200\mev$.
Further, it is apparent that $c\tau$ will be determined to better
than $\delta c\tau\sim 0.4$ mm at the $1\sigma$ level, corresponding
roughly to a $\pm 2\%$ measurement of $c\tau$.

\begin{figure}[!ht]
\leavevmode
\begin{center}
\epsfxsize=5in
\hspace{0in}\epsffile{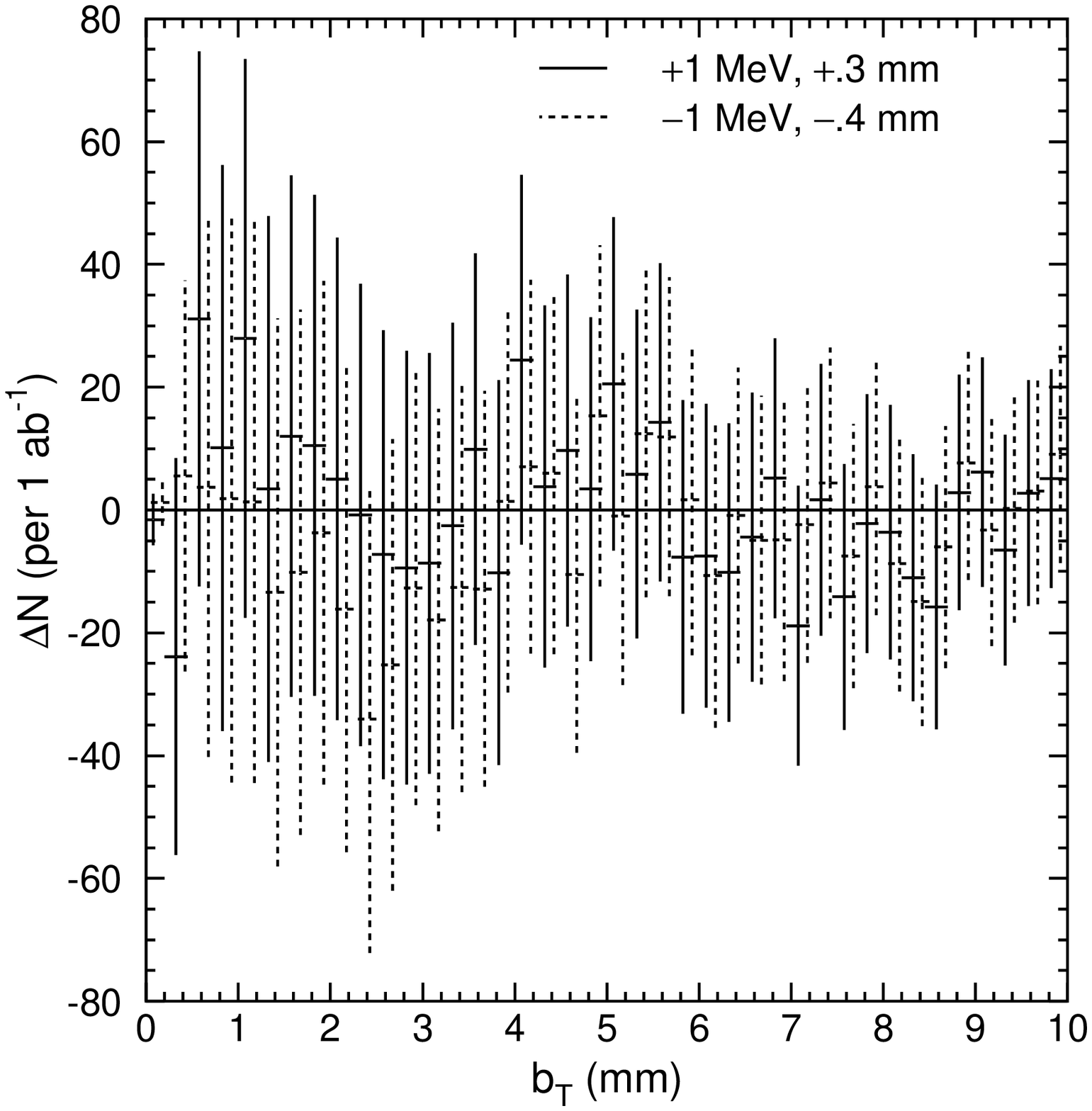} 
  \caption{For $\mcpmone=175\gev$ and $\rts=600\gev$,
we plot the differences, $\Delta N$, (of the number of entries
in a given bin) between the $b_T$ spectra for the central values
of $[\dmchi=200\mev,c\tau=23.81~\mbox{mm}]$ and those for 
$[\dmchi=201\mev,c\tau=24.11~\mbox{mm}]$ and 
$[\dmchi=199\mev,c\tau=23.41~\mbox{mm}]$. 
These differences plotted are those found after choosing the 
normalizations of the distributions for
the latter two alternative parameter choices so as to minimize $\Delta\chi^2$.
Also shown are the errors for a total luminosity of $L=1\abi$.
We have artificially shifted the $b_T$ values at which the points
are plotted so that the two different cases can be distinguished.
}
\label{btspectrum_diffs}
\end{center}
\end{figure}

It is perhaps useful to see what features of the $b_T$
distribution allow such precise parameter determination. To this end,
we plot in Fig.~\ref{btspectrum_diffs} the differences, $\Delta N$,  in the
number of entries in a given $b_T$ bin between the distributions for
the central $[\dmchi=200\mev,c\tau=23.81~\mbox{mm}]$ parameter choice
and those for 
$[\dmchi=201\mev,c\tau=24.11~\mbox{mm}]$ and 
$[\dmchi=199\mev,c\tau=23.41~\mbox{mm}]$. The $c\tau$ choices
for these latter two $\dmchi$ values are those which minimize
the $\Delta\chi^2$, as shown in Fig.~\ref{ctaudmchisq}.
The $\Delta N$ values are those obtained after choosing the relative
normalizations of the alternative parameter distributions
so as to minimize the $\Delta\chi^2$. 
Also shown are the errors
for a total luminosity of $L=1\abi$. We observe a systematic
variation, especially at low $b_T$, of the $\dmchi=201\mev$
and $\dmchi=199\mev$ distributions relative to that for
$\dmchi=200\mev$. There are a
large number of bins, each with $\Delta\chi^2\gsim 0.5$, which combine
to give the significant overall $\Delta\chi^2$ plotted in
Fig.~\ref{ctaudmchisq}. (We have checked that these differences
between the 
distributions are not an artifact of the Monte Carlo simulation statistics.)
Note for instance the depletion in the smallest $b_T$ bin for
$\dmchi=201\mev$, mentioned earlier, followed rapidly by an excess in
the next set of $b_T$ bins.
We hope that these distributions make it apparent that both can
be distinguished at a good level from that for
$[\dmchi=200\mev,c\tau=23.81~\mbox{mm}]$ and that the distributions for 
the two alternative parameter choices can also be clearly
distinguished from one another.

In the above analysis, we have implicitly assumed
that there are no contaminating background events.
After our minimal $p_T^\gam>10\gev$ tag cut, 
backgrounds (including the two-photon backgrounds) 
are unlikely to yield events having significant
impact parameters.\footnote{If $c\tau$ is of the order of $100~\mu$m,
then a precise accounting for the $\epem\to\epem\gam\tau^+\tau^-\to\gam
\pi^+\pi^-\emiss$ background might be necessary.}

A measurement of $c\tau$ can be converted to a joint constraint on
$\dmchi$ and  the underlying $M_1,M_2,\mu,\tanb$ parameters.
Consider $\dmchi<700\mev$ for which the only modes
of any importance are $\cpmone\to\ell^\pm\nu\cnone$ and $\cpmone\to
\pi^\pm\cnone$. For the latter, we have
\bea
\Gamma(\tilde{\chi}_1^- \rightarrow \tilde{\chi}_1^0 \pi^-)
&=& {f_\pi^2 G_F^2\over 4 \pi} \frac {|\vec{k}_\pi|}{\mcpmone^2}
\left\{ \left( O^L_{11} + O^R_{11} \right)^2 \left[ \left(
\mcpmone^2 - \mcnone^2 \right)^2 - m^2_\pi \left(
\mcpmone - \mcnone\right)^2 \right]
\right. \nonumber \\ &\phantom{=}& \left.
\hspace*{25mm} + \left( O^L_{11} - O^R_{11} \right)^2 \left[ \left(
\mcpmone^2 - \mcnone^2 \right)^2 - m^2_\pi \left(
\mcpmone + \mcnone \right)^2 \right] \right\}\,,
\label{pidecay}
\eea
where the $O_{11}^{L,R}$ describe the $\wmp\cpmone\cnone$ coupling,
\beq
O_{11}^L= -{1\over\sqrt 2}N_{14}V_{12}+N_{12}V_{11}\,,\quad
O_{11}^R= +{1\over\sqrt 2}N_{13}U_{12}+N_{12}U_{11}\,.
\label{oforms}
\eeq
Current experiment gives $f_\pi=92.42\pm 0.3\mev$, which yields a $\sim 0.7\%$
error for this width. More important is the parameter
dependence of $O_{11}^{L,R}$. Naively, these are $\sim 1$ for the
wino-like LSP scenario. However, their dependence on the
underlying SUSY parameters $M_1,M_2,\mu,\tanb$ is significant.
For example, for $\mcpmone\sim 175\gev$, they vary
by about $5\%$ as $|\mu|$ increases in the range $500\gev$ to $1\tev$,
with much more dramatic variation as $|\mu|$ approaches $300\gev$.
At large $|\mu|>500\gev$ the $O_{11}^{L,R}$ also vary by about $(1-2)\%$
as $\tanb$ ranges from small to large values.  At smaller $|\mu|$,
the $\tanb$ variation is much more dramatic.
The widths for the $\cpmone\to\ell^\pm\nu\cnone$ ($\ell=e,\mu$) modes 
are also proportional to combinations of the $O_{11}^{L,R}$. 
If the $O_{11}^{L,R}$ are held fixed at the Ref.~\cite{cdg1} central values, 
Eq.~(\ref{pidecay}) implies that a shift in $\dmchi$ by $1\mev$ corresponds
(in the vicinity of $\dmchi\sim 200\mev$) to a shift in $c\tau$
by about 2.5\%. However, there are many parameter
choices that yield the same $\dmchi$ with $O_{11}^{L,R}$ values that differ
by more than this. Given these uncertainties and the
$\sim 2\%$ experimental error for the $c\tau$ determination,
it will be difficult to constrain $\dmchi$
more accurately than via the experimentally direct kinematic distribution
and $b_T$ shape fit techniques.  Ultimately,
the reverse strategy might prove useful. That is, determine $\dmchi$
directly from the kinematic distributions and $b_T$ distribution shape
and $c\tau$ from the $b_T$ distribution shape and then use these values
to constrain 
the $M_1,M_2,\mu,\tanb$ parameters. Note that it would be crucial to include
the one-loop corrections to $\dmchi$ in this process.

\noindent{\bf \boldmath 5. Determining $M_2$, $\mu$ and $\tanb$.}
Let us suppose that we have accurate determinations of $\mcpmone$
and $\dmchi$ using the kinematic distribution techniques described above. 
We will then wish to extract the underlying SUSY parameters.
We study our ability to do so for the specific case of
$\mcpmone=175\gev$ and $\dmchi=200\mev$. 
This particular case is motivated in the context
of ($\delgs=0$ O-II)/AMSB one-loop
boundary conditions: (a) $\mcpmone<200\gev$ 
is preferred if we are to avoid the extreme fine-tuning
for larger $\mcpmone$ implied by the large value ($\sim 10$)
predicted for $M_3/M_2$; and (b) 
$\dmchi<200-500\mev$ is preferred  because
of the large value predicted for $M_1/M_2$. 
We again remind the reader that,
for such $\dmchi$, the $\gam\cpone\cmone$
events we employ will be background free since 
at least one of the final $\pi$'s will have an observable HIP.

\begin{figure}[!ht]
\leavevmode
\begin{center}
\epsfxsize=6.25in
\hspace{0in}\epsffile{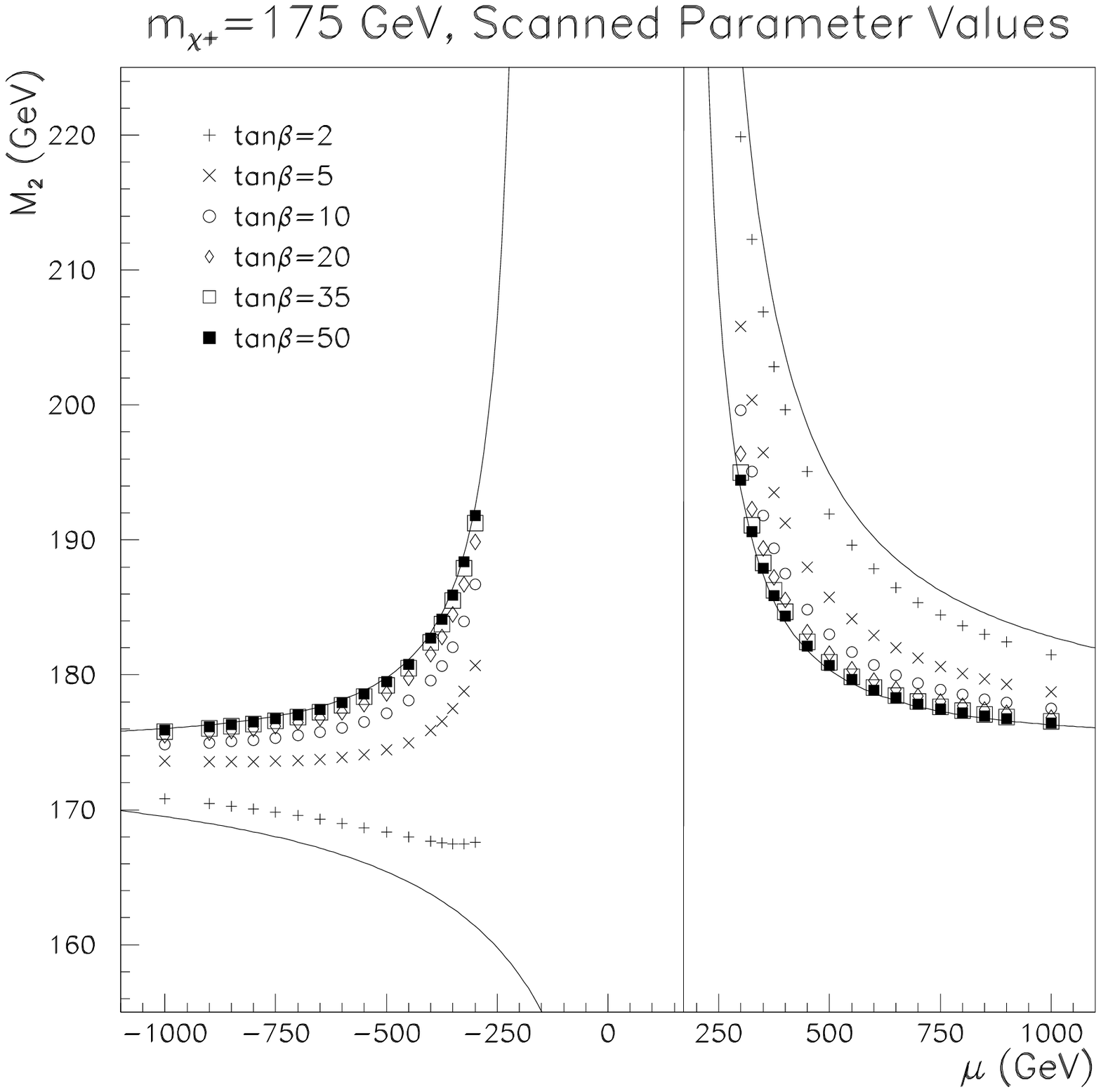} 
  \caption{For $\mcpmone=175\gev$, the outer solid lines show the 
region in the $[\mu,M_2]$ plane consistent with $1\leq\tanb\leq 100$
and $|\mu|>150\gev$. Also shown are the values of
$[\mu,M_2]$ sampled for $\tanb=2,5,10,20,35,50$ in determining
our fits to the $\epem\to\gam\cpone\cmone$ cross sections.
}
\label{scanregion}
\end{center}
\end{figure}

The SUSY parameters 
entering into the chargino sector at tree level are $M_2$, $\mu$ and $\tanb$.
The measured value of $\mcpmone$ provides only one constraint
on these three parameters.  The projection of this constraint
on the standard $[\mu,M_2]$ plane gives the two regions indicated
in Fig.~\ref{scanregion} by the outer solid lines.
In this paper, we restrict
ourselves to $|\mu|>300\gev$, since for lower $|\mu|$
the $\cpmtwo$ charginos would also be produced and the nature
of the analysis and parameter extraction procedures would change dramatically.
Typical radiative electroweak symmetry breaking favors large $|\mu|$ values.

The goal is to use the $\gam\cpone\cmone$ production cross
section and kinematic dependencies to obtain
additional constraints on the three parameters. Sensitivity
to these parameters arises entirely through the
coupling of the $Z$ to $\cpone\cmone$:
\beq
{ig\over \cw}\gamma^\mu\left(O_{11}^{\prime\,L} P_L+O_{11}^{\prime\,R}
P_R\right)\,,\quad
\hbox{with}~
O_{11}^{\prime\,L}=-\cw^2+{1\over 2}V_{12}^2\,,\quad 
O_{11}^{\prime\,R}=-\cw^2+{1\over 2}U_{12}^2\,,
\label{couplings}
\eeq
where $\cw\equiv \cos\theta_W$ etc., and $V_{12}$ and $U_{12}$ are 
elements of the matrices that diagonalize the chargino mass
matrix. (In our work, we neglect possible CP-violating phases
in the chargino sector.) Their squares can be written in the form
$V_{12}^2={1\over 2}(1-\cos2\phi_L)$ and 
$U_{12}^2={1\over 2}(1-\cos2\phi_R)$ with
\bea
\cos2\phi_L&=&- {M_2^2-\mu^2-2\mw^2\cos2\beta\over
\sqrt{(M_2^2+\mu^2+2\mw^2)^2-4(M_2\mu-\mw^2\sin 2\beta)^2}}\,,
\nonumber\\
\cos2\phi_R&=&- {M_2^2-\mu^2+2\mw^2\cos2\beta\over
\sqrt{(M_2^2+\mu^2+2\mw^2)^2-4(M_2\mu-\mw^2\sin 2\beta)^2}}\,.
\label{cosphidefs}
\eea
Asymptotic expressions for $V_{12}$ and $U_{12}$, valid for
$|M_2\pm\mu|\gg \mz$ are \cite{ghino} (using $\cb\equiv\cos\beta$ etc.):
\beq
V_{12}={\mw\sqrt2(M_2\sb+\mu\cb)\over M_2^2-\mu^2}\,,\quad
U_{12}={\mw\sqrt2(M_2\cb+\mu\sb)\over M_2^2-\mu^2}\,.
\label{uvforms}
\eeq
Clearly $V_{12}^2$
and $U_{12}^2$ are small compared to $\cw^2$ when $|\mu|$ is large.
Further, $U_{12}^2\gg V_{12}^2$ is typical for $\tanb>2$ when
$|\mu|\gg M_2$.
In the scenarios we consider, the $\cpmone$ are indeed highly wino-like
and, thus,  the problem is to pick out the small $V_{12}^2$ and $U_{12}^2$
corrections to the dominant $-c_W^2$ term in the $Z$ coupling.

To understand how to proceed,
it is useful to first briefly review results
for $\epem\to\cpone\cmone$ without a photon tag.
Following \cite{choi}, we write
\beq
T(\epem\to\cpone\cmone)={e^2\over s}\sum_{\alpha,\beta=R,L}
Q_{\alpha\beta}(s)\left[\anti v(e^+)
\gamma_\mu P_\alpha u(e^-)\right]\left[\anti u(\cmone)\gamma^\mu P_\beta
v(\cpone)\right]
\eeq
where
\bea
Q_{LL}(s)&=&1+{D_Z(s)\over s_W^2c_W^2}(s_W^2-{1\over 2})\left(-c_W^2+{1\over 2}
V_{12}^2\right)\,,
\nonumber\cr
Q_{LR}(s)&=&1+{D_Z(s)\over s_W^2c_W^2}(s_W^2-{1\over 2})\left(-c_W^2+{1\over 2}
U_{12}^2\right)+{D_{\snu}(s,t)\over 4s_W^2}(2-U_{12}^2)\,,
\nonumber\cr
Q_{RL}(s)&=&1+{D_Z(s)\over c_W^2}\left(-c_W^2+{1\over 2}V_{12}^2\right)\,,
\nonumber\cr
Q_{RR}(s)&=&1+{D_Z(s)\over c_W^2}\left(-c_W^2+{1\over 2}U_{12}^2\right)\,,
\label{qdefs}
\eea
In Eq.~(\ref{qdefs}), $D_Z(s)=s/(s-\mz^2)$ and $D_{\snu}(s,t)=s/(t-\msnu^2)$. 
It is crucial to observe that
$Q_{RL}\sim {1\over 2 c_W^2}V_{12}^2$ and 
$Q_{RR}\sim {1\over 2 c_W^2}U_{12}^2$ at large $s$, implying that
a right-handed polarized $e^-$ beam will provide a very direct probe of
$V_{12}^2$ and $U_{12}^2$, but at the sacrifice of a very suppressed
cross section. Pure $e_R^-$ also has the advantage of eliminating
the `background' from the $\snu_e$ exchange diagram. Since
this diagram will significantly suppress the unpolarized cross section
for $m_{\snue}<1\tev$ (i.e. even for masses for which we will not be able
to directly detect the $\snue$ and measure its mass),
the normalization of the unpolarized cross
section cannot be calculated reliably enough to use 
as an ingredient in extracting the
parameters of the model. The importance of the polarized measurements,
for which this is not a problem, implies that
maximal integrated luminosity will thus be of paramount importance, whether
one is looking at an mSUGRA-type SUSY breaking scenario or, as we shall discuss
in detail shortly, the degenerate $\mcpmone\sim\mcnone$ scenario.
In terms of the $Q_{\alpha\beta}$'s, the $\epem\to \cpone\cmone$
amplitude squared is
\beq
|{\cal A}|^2\propto 
 ( - 2 \pc\cdot\pem \pcp\cdot\pep s (\qlr^2 + \qrl^2) - 2 \pc\cdot\pep \pcp\cdot\pem s (\qll^2 +
 \qrr^2) - s^2 \mcpmone^2 (\qll \qlr + \qrl \qrr))/2
\label{ampsq}
\eeq
where  $p_{\cmone}\cdot p_{e^-}=p_{\cpone}\cdot p_{e^+} 
\simeq E_b^2(1-\beta \cos\theta)$, 
$p_{\cmone}\cdot p_{e^+}=p_{\cpone}\cdot p_{e^-}
\simeq E_b^2(1+\beta\cos\theta)$, $\beta^2=1-\mcpmone^2/E_b^2$
and $E_b=\rts/2$.
By measuring the magnitude of the cross section,
especially for a right-handed polarized $e^-$ beam, and its dependence
on $\cos\theta$, one can hope to determine the charges $Q_{\alpha\beta}$
and, thence, the crucial $V_{12}^2$ and $U_{12}^2$ matrix entries 
that probe $\mu$, $M_2$ and $\tanb$. The $\cpmone$ are not
directly observed (unless they have a substantial path length,
as is possible in this model but not assumed in our analysis),
but, by using the accurately
measured values of $\mcpmone$ and $\dmchi$ and the measured
three-momenta of the $\pi^\pm$, the chargino momenta can be reconstructed
up to a two-fold ambiguity. The analogue of
this reconstruction for the $\gam\cpone\cmone$ final state is discussed
in the Appendix.
We note that $|{\cal A}|^2$ and, thence, the
integrated cross section for any particular machine energy and set of cuts
depend bi-quadratically on $V_{12}^2$ and $U_{12}^2$.

\begin{figure}[!ht]
\leavevmode
\begin{center}
\epsfxsize=6.25in
\hspace{0in}\epsffile{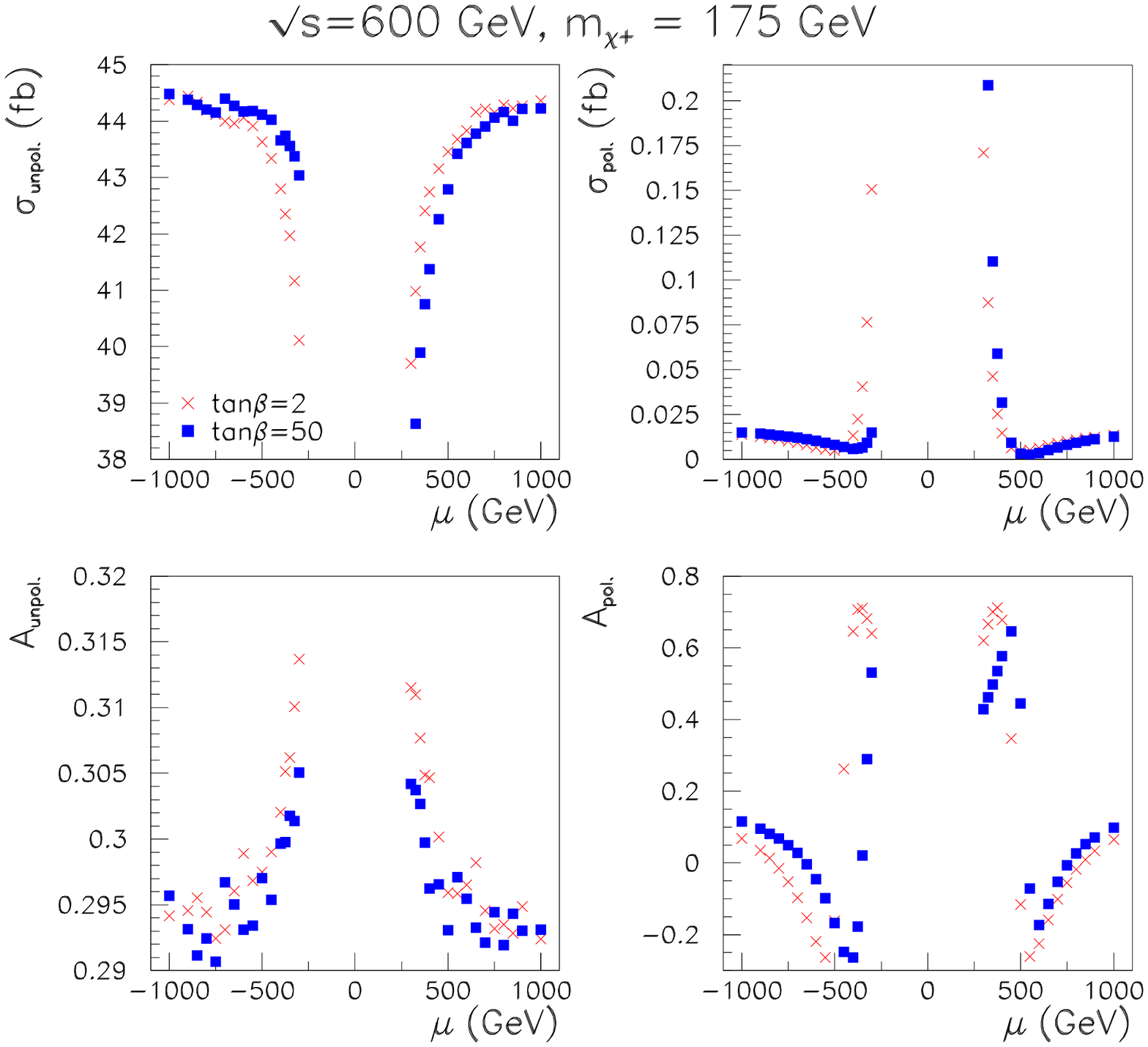}
  \caption{For $\mcpmone=175\gev$, we plot $\sigma$ and $A$
as a function of $\mu$ for $\tanb=2$ and $50$.
Results are given for an unpolarized $e^-$ 
beam and a purely $e_R^-$ beam at $\rts=600\gev$.
A mass splitting of $\dmchi=200\mev$ was employed in generating the $\pi^\pm$
momenta.
}
\label{xscnsasyms}
\end{center}
\end{figure}

 These same `charges', $Q_{\alpha\beta}$, appear in 
the expression for the $\epem\to\gam
\cpone\cmone$ amplitude, which contains
two initial state radiation diagrams 
summed over $Q_{\alpha\beta}(q^2=\mchichi^2)$'s
and two final state radiation diagrams summed over $Q_{\alpha\beta}(q^2=s)$'s,
where $q$ is the four-momentum carried by the virtual $Z$ or $\gam$.
The resulting form for $|{\cal A}|^2$  has a complicated
dependence on many kinematical variables. (In the Appendix,
we give the structure of $|{\cal A}|^2$ for the dominant
initial state radiation diagrams. An expression
in a different formalism for the full $|{\cal A}|^2$
is given in \cite{Datta:1999yw}.) In the soft-photon
limit, $|{\cal A}(\epem\to\gam\cpone\cmone)|^2$
is directly proportional to $|{\cal A}(\epem\to \cpone\cmone)|^2$,
but this limit does not allow an understanding of all of the
features of the $\gam\cpone\cmone$ cross section. However,
it remains true that
right handed electron polarization produces a very suppressed
cross section with high sensitivity to $V_{12}^2$ and $U_{12}^2$ and no
sensitivity to the probably unknown $\snue$ mass.

We now outline our precise procedures. For each event, the observed
$\gam$ defines a chargino-pair, i.e. $\mchichi$, center of mass system.  
In this c.m.s., we have followed the Collins-Soper procedure \cite{colsop} of
defining a $z$-axis by boosting along the $\gam$ direction to the $\mchichi$
c.m.s., determining the unit vectors $\what u_{e^\pm}$ in the $e^{\pm}$
three momentum directions after the boost and defining
$\what z=(\what u_{e^+}-\what u_{e^-})/(2-2\what u_{e^+}\cdot\what u_{e^-})$.
For the vast majority of events, $\what z$ is very closely aligned with
the $z$ axis in the laboratory frame. We then determine the $\cpmone$
momenta in this frame (using the observed
$\pi^\pm$ momenta and the known values of $\mcpmone$
and $\mcnone=\mcpmone-\dmchi$) following the procedure of the Appendix
and compute the angle, $\what\theta$, of $\vec p_{\cmone}$
with respect to $\what z$. ($\vec p_{\cpone}=-\vec p_{\cmone}$
in the $\mchichi$ c.m.s.)  Since there
is a two-fold ambiguity in the reconstruction, we bin events twice
corresponding to the two possible reconstructed $\cpmone$ three-momenta.

We have restricted our analysis to the absolute magnitude
of the cross section, $\sigma$, and the asymmetry
\beq
A\equiv {\sigma(\cos\what\theta>0)-\sigma(\cos\what\theta<0)\over
\sigma(\cos\what\theta>0)+\sigma(\cos\what\theta<0)}
\eeq
where the cross sections are obtained by integrating
over all of phase space consistent 
with the specified sign of of $\cos\what\theta$ 
and our basic $p_T^\gamma>10\gev$ and $10^\circ<\theta_\gamma<170^\circ$
photon-tag cuts. We illustrate the dependence of $\sigma$ and $A$
on $\mu$ for $\tanb=2$ and $50$ in Fig.~\ref{xscnsasyms}.
($M_2$ is determined as a function of $\mu$ at a given
$\tanb$ by the fixed value of $\mcpmone=175\gev$ as shown in
Fig.~\ref{scanregion}.) Results are given for unpolarized beams
and for the case of a purely right-handed polarized electron beam.
For $L=1\abi$, typical $1\sigma$ errors for $\sigma$ and $A$ are
of order $0.5\%$ and $1\%$ ($15\%$ and $50\%$) for
the unpolarized (polarized) cases, respectively.
Relative to the above-quoted typical errors,
there is significant variation with respect to $\mu$ of both $\sigma$
and $A$, but different $\tanb$ values will be difficult to separate
since a large change in $\tanb$ can be compensated by a small shift in $\mu$ 
that will leave the four observables more or less unchanged.  
The variations with $\mu$ are especially
dramatic for $\sigma_{\rm polarized}$ and $A_{\rm polarized}$,
but the large statistical errors even for $L=1\abi$ render these variations
no more (or less) useful than those for the corresponding
observables in the unpolarized beam case. Since the polarized
and unpolarized cross sections and asymmetries contain different weightings
of $U_{12}^2$ and $V_{12}^2$, accumulating significant
luminosity in both modes is useful for maximizing our ability to
determine $U_{12}^2$ and $V_{12}^2$.

\begin{table}[h!]
\renewcommand{\arraystretch}{1}
  \begin{center}
    \begin{tabular}[c]{|c|c|c|c|c|c|c|} \hline
\ & $a$ & $b$ & $c$ & $d$ & $e$ & $f$ \\
\hline
$\xup$ &       28.7596 &  
      -5.48626 & 
      -15.5788 & 
      0.20296 & 
      1.07641 & 
      5.98101 \\
$\xum$ &   15.7698 & 
      -7.08801 & 
      -4.67496 & 
      0.0723394 & 
      -1.1899 & 
      4.98557 \\
$\xpp$ &       0.0111671 & 
      -0.323629 & 
      -0.124093 & 
      2.91672 & 
      0.637143 & 
      1.26579 \\
$\xpm$ &       0.00771917 & 
      -0.103494 & 
      -0.169943 & 
      0.59423 & 
      1.27397 & 
      0.784927 \\
\hline
%
    \end{tabular}
    \caption{Tabulation of fitted parameters $a,b,c,d,e,f$
(in units of fb's) of Eq.~(\protect\ref{sigform})
for each of the four cross sections employed in our analysis.
Results are for $\rts=600\gev$
and are those obtained after Monte Carlo 
integration incorporating cuts and the chargino reconstruction
algorithm described.}
    \label{abcdef}
  \end{center}
\end{table}

To simplify our analysis, we scanned a large selection of
$\mu$ values in the ranges $[-1000\gev,-300\gev]$
and $[300\gev,1000\gev]$ at a series of $\tanb$ values
($\tanb=2,5,10,20,35,50$)\footnote{The value of $M_2$ at each point is
determined by $\mcpmone$.}  and fit the results for
1) $\xup\equiv\sigma_{\rm unpolarized}(\cos\what\theta>0)$,
2) $\xum\equiv\sigma_{\rm unpolarized}(\cos\what\theta<0)$, 
3) $\xpp\equiv\sigma_{\rm polarized}(\cos\what\theta>0)$ and
4) $\xpm\equiv\sigma_{\rm polarized}(\cos\what\theta<0)$
to the  bi-quadratic form (required by the amplitude structure)
\beq
\sigma_i=a+b\usq+c\vsq+d(\usq)^2+e(\vsq)^2+f\usq\vsq\,.
\label{sigform}
\eeq
Results for $a-f$ for $\rts=600\gev$, $\mcpmone=175\gev$
and $\dmchi=200\mev$ (the latter possibly affecting  
reconstruction of the angle of the chargino) are given  
in Table \ref{abcdef}.
These fits give the correct cross section with an accuracy of
better than 0.7\% for all the sampled points.
For $\xpp$ and $\xpm$, note the very small constant terms  and
the much larger quartic term coefficients ($d,e,f$).
As already discussed, this is to be expected and follows from
the wino-like nature of the $\cpmone$.

\begin{figure}[!ht]
\leavevmode
\begin{center}
\epsfxsize=6.25in
\hspace{0in}\epsffile{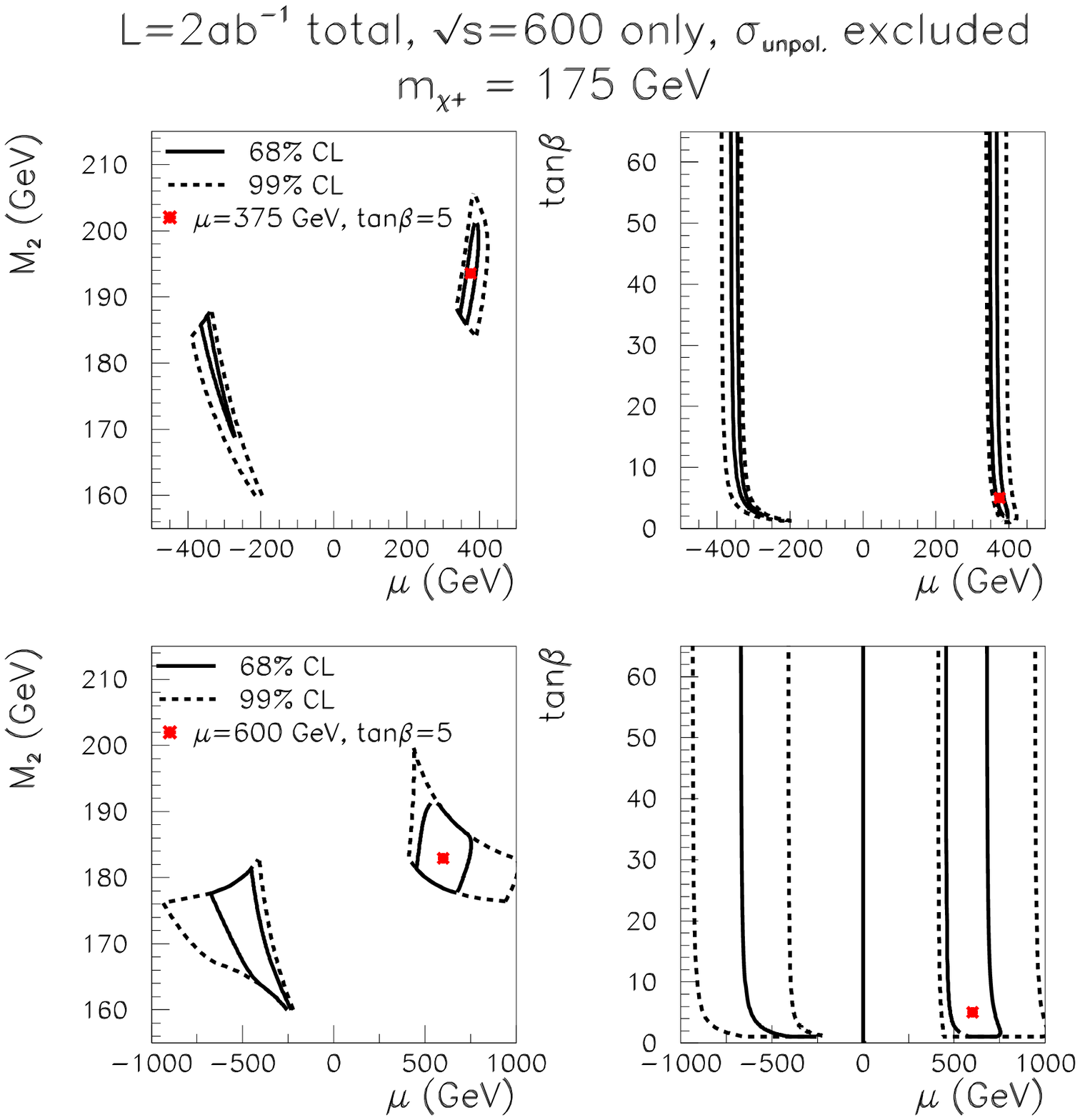}
  \caption{For $\mcpmone=175\gev$, we plot 
the contours for $\Delta\chi^2=2.3$ (68\% CL for two independent
variables) and $\Delta\chi^2=13$ (99\% CL) in the
$[\mu,M_2]$ and $[\mu,\tanb]$ parameter spaces
using the two reference cases
of $[\tanb,\mu]=[5,375\gev]$ and $[5,600\gev]$.
We have assumed running at $\rts=600\gev$ sufficient to accumulate
integrated luminosities of $L=1\abi$ for both unpolarized beams
and for a purely right-handed electron beam.
The $\Delta\chi^2$ has been computed using data for
$\sigma_{\rm polarized}$, $A_{\rm unpolarized}$
and $A_{\rm polarized}$; $\sigma_{\rm unpolarized}$ was not included.
A mass splitting of $\dmchi=200\mev$ was used in generating the pion momenta.}
\label{scan600}
\end{center}
\end{figure}

\begin{figure}[!ht]
\leavevmode
\begin{center}
\epsfxsize=6.25in
\hspace{0in}\epsffile{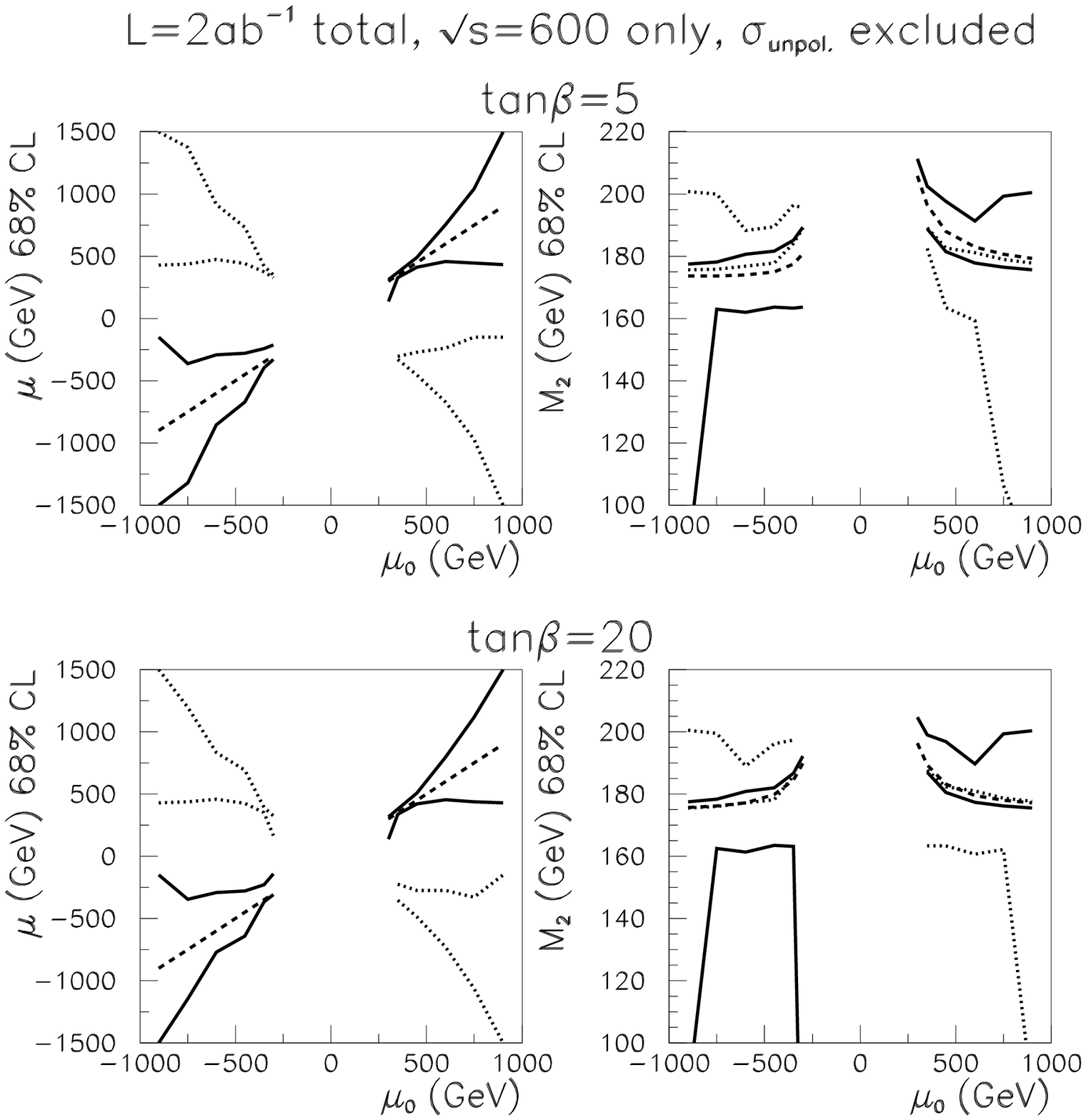}
  \caption{For $\mcpmone=175\gev$, we plot the upper and lower 
$\Delta\chi^2=2.3$ (68\% CL) limits
for $\mu$ and $M_2$ as a function of the
input value of $\mu$, $\mu_0$, for input values of $\tanb=5$ and $\tanb=20$.
The dashed lines indicate either the input values of $\mu_0$ or the
corresponding $M_2(\mu_0)$ values for the input $\tanb$, given $\mcpmone=175\gev$.
For each $\mu_0$, the 68\% CL boundaries for choices of $\mu$ with the same 
(opposite) sign of $\mu$ are shown by solid (dotted) lines.
Points not plotted at small $\mu\sim 300\gev$ are inconsistent
with the given input $\mcpmone=175\gev$.
Data inputs are as specified in Fig.~\ref{scan600}. } 
\label{clerrors}
\end{center}
\end{figure}

We now employ these cross section fits to study the accuracy
with which we can expect to determine $M_2$, $\mu$
and $\tanb$. To illustrate, we give
in Fig.~\ref{scan600}, for $\rts=600\gev$,
contours of $\Delta\chi^2=2.30$ (68\% CL for two independent parameters)
and $\Delta\chi^2=13$ (99\% CL),
using reference models of $[\tanb,\mu]=[5,375\gev]$
and $[5,600\gev]$. For this figure, we employ the following
set of observables: the forward/backward asymmetry for
unpolarized beams; the absolute cross section for polarized beams;
and the forward/backward asymmetry for polarized beams.\footnote{We
compute the error in $A\equiv {N^+-N^-\over N}$
as $\delta A=\sqrt{{1-A^2\over N}}$, where $N=N^++N^-$.}
As explained later, the absolute unpolarized cross section
is likely to have systematic theoretical
uncertainties that are much larger than the statistical errors,
and is not employed.
We observe that, at the 68\% CL, 
$M_2$ and $|\mu|$ can be determined to $\sim 8\%$ and $\sim 16\%$,
respectively, for $|\mu|\sim 400\gev$ and to $\sim 8\%$ and $\sim 40\%$,
respectively, for $|\mu|\sim 600\gev$;
$\tanb$ is essentially undetermined. The 68\% CL limits
on $\mu$ and $M_2$ for a wider selection of input $\mu$
values and input $\tanb$ values of 5 and 20 are shown in
Fig.~\ref{clerrors}. Results for other $\tanb$ values
are quite similar. Unless $\tanb$ has been determined
by other data, we must combine all such graphs
and take the outer errors. Clearly, at the highest $|\mu|$ values, errors 
for the $\mu$ and $M_2$ determinations become quite large, and
significant uncertainty in $M_2$ develops as $\mu$ falls below $300\gev$.

The unpolarized cross section can provide useful constraints on SUSY
parameters only if
systematic errors are substantially smaller than the typical
5\%-10\% variation of $\sigma_{\rm unpolarized}$ as a function of parameters;
see Fig.~\ref{xscnsasyms}.
Systematic {\it experimental} errors at an electron collider 
should be quite small, probably even smaller
than the $\sim 0.5\%$ statistical errors.
However, theoretical uncertainties in $\sigma_{\rm unpolarized}$
must also be well below the 5\%-10\% level.
Even if we assume that
the $\snue$ contribution can be absolutely normalized 
(for example, using $\snue$ observations from the LHC) 
higher-order electroweak corrections to the $\epem\to
\gam\cpone\cmone$ cross section will need to be computed. In addition,
one will be implicitly assuming that the supersymmetric $Z\cpone\cmone$
coupling strength is indeed precisely
that predicted by assuming strict supersymmetry
for tree-level couplings. This will be difficult to independently check. 
Finally, to employ the absolute normalization of the unpolarized cross section
one would need to rely on the modeling employed for computing the
$\cpmone\to\pi^\pm\cnone$ branching ratio at the $\lsim 5\%$ level.
The computation for the $\cpmone\to\ell^\pm\nu\cnone$ and
$\cpmone\to\pi^\pm\cnone$ widths was reviewed in the previous section.
While some of the theoretical uncertainties in the widths cancel
when computing the branching ratios, the latter will still have
dependence on the $M_1,M_2,\mu,\tanb$ parameters at the $(2-5)\%$
level for $\dmchi\lsim 700\mev$ (much larger as the multi-pion
decay modes of the $\cpmone$ enter at somewhat larger $\dmchi$).
For this paper, we have assumed that the absolute normalization of the
unpolarized absolute cross section is not computable to the 5\% level,
and have determined the $\Delta\chi^2$ values using only $A_{\rm unpolarized}$,
$\sigma_{\rm polarized}$ and $A_{\rm polarized}$. 
If we were able to reliably use $\sigma_{\rm unpolarized}$,
the resulting plots analogous
to Fig.~\ref{scan600} would show a moderate increase in our ability
to determine $M_2$ and $\mu$, but $\tanb$ would remain 
essentially undetermined.

We have also examined whether it is useful 
to include data from several $\rts$ choices for determining
$\mu,M_2,\tanb$. We found that splitting the total luminosity
between $\rts=450\gev$ and $600\gev$ 
actually causes a mild decrease in our ability to determine
the parameters (assuming no large variation
of instantaneous luminosity with $\rts$).  
This remains true even if we run at both energies
and use as an observable the ratio $\sigma_{\rm unpolarized}(\rts=450\gev)/
\sigma_{\rm unpolarized}(\rts=600\gev)$ (which would 
be less affected by systematic theoretical uncertainties
as compared to the individual absolute 
$\sigma_{\rm unpolarized}$ normalizations).
If we are conservative and rely only on polarized data
($L=2\abi$ at $\rts=600\gev$), the $M_2$ and $\mu$
accuracies are almost the same as shown in Fig.~\ref{scan600}.

Additional information can, in principal, be gleaned from the 
$\cpmone\to\cnone\pi^{\pm}$ decays. However, this is quite
difficult in practice.  The issue is whether one can use
correlations involving the pions in the final state. Correlations
between the pions themselves or between the pions
and the reconstructed momenta for the $\cpmone$ will be present
to the extent that information regarding the direction
of the spins $s^\pm$ of the decaying $\cpmone$ influences the $\pi^\pm$.
Consider $\cmone$ decay. For a given $s^-$ choice, one finds
\beq
|{\cal M}(\cmone\to \pi^-\cnone)|^2\propto 1+{\dmchi-p_{\pi^-}\cdot s^-
\over \mcpmone}
\,
\eeq
after dropping terms higher order in the small ratios 
$\dmchi/\mcpmone$ and $p_{\pi^-}/\mcpmone$. Since $|\vec p_{\pi^-}|$
is of order $\dmchi$,
the dependence on $p_{\pi^-}\cdot s^-$ is obscured
by the much larger leading term. For this reason, we
have not attempted to employ these correlations.

\noindent{\bf 6. Discussion and Conclusions.}
We have considered the detection at a linear $\epem$ collider
of the lightest chargino in a model in which it is closely
degenerate with a wino-like neutralino.
The phenomenology depends critically upon
the mass splitting $\dmchi\equiv \mcpmone-\mcnone$. For $\dmchi<m_\pi$,
one will observe
long-lived heavily ionizing $\cpmone$ tracks. For $\dmchi\gsim 2\gev$,
the usual mSUGRA jets + missing energy signal will be dominant.
Both can be detected for $\mcpmone$ essentially all the way up to
the threshold at $\rts/2$ assuming modest integrated luminosity of
$L=50\fbi$. For $\mcpmone$ not near threshold, much less luminosity
would be adequate.
For the range $200\mev<\dmchi<1\gev$, favored in typical models, 
the $\gam+\pi^+\pi^- +\emiss$ final state will be the crucial discovery mode.
Then, so long as soft $\gsim 200\mev$
pions are visible in the detector,\footnote{If the soft
pions are not visible, $\gam\cpone\cmone$ production yields a $\gam+
~{/ \hskip-5pt E}$ final state with substantial background from
$\gam+Z(\to\nu\anti\nu)$, and discovery reach will be
more limited and the error for $\mcpmone$, summarized shortly, much larger;
$\dmchi$ will not be measurable.}
 discovery in the $\gamma+\pi^+\pi^-+\emiss$ mode will
be possible up to $\mcpmone\sim (\rts-p_T^\gam)/2$ for $L=50\fbi$. 
In particular, we argued that simple cuts requiring $p_T^\gamma>10\gev$, 
only two soft/central pions, large (mainly invisible) mass
recoiling opposite the trigger $\gam$, and no small angle ($\theta>1^\circ$)
$e^-$ or $e^+$ (as would be present for the
most obvious two-photon backgrounds) have an excellent chance of
suppressing backgrounds to a negligible level.
An overview of the discovery modes relevant as a function of location
in the full $[\mcpmone,\dmchi]$
parameter space appeared in Fig.~\ref{regions}.

We then turned to a more detailed study of scenarios that would
arise in the $\delgs=0$ O-II string model and in the AMSB model.
These models predict the same low-energy
ratios for the gaugino masses, $M_2:M_1:M_3\sim 0.3:1:3$ (tree-level),
and generally require substantial $|\mu|$ for automatic
electroweak symmetry breaking. In these models, the smallness of $M_2$
compared to the other mass parameters implies that  
$\dmchi$ will almost certainly lie in the $\mpi<\dmchi<800\mev$
range (after including radiative corrections) for which the only
appropriate mode for discovery and study is
 $\epem\to\gam\cpone\cmone\to \gam\pi^+\pi^-\cnone\cnone$.
The large ratio of $M_3/M_2$ implies that fine tuning for these models
will be quite extreme unless 
the chargino mass lies in the $<200\gev$ range. 
For $\mcpmone$ and $\dmchi$ in these ranges we studied  
the accuracy with which $\mcpmone$ and $\dmchi$ can be measured
using kinematic distributions of the pions in the final state.
For the particularly typical choices of 
$\mcpmone=175\gev$ and $\dmchi=200\mev$,
we studied the accuracy with which the $\dmchi$ and $c\tau$ could
be determined from the transverse impact parameter distribution
for the soft pion tracks. Finally, for $\mcpmone=175\gev$ and
$\dmchi=200\mev$, we determined the precision with which  
the fundamental underlying SUSY parameters
$M_2$, $\mu$ and $\tanb$ could be determined from
cross sections and asymmetries. We summarize the results of these 
studies below.

The errors estimated for $\mcpmone$ and $\dmchi$
assume that the electron sneutrino is not so light as to substantially
suppress the unpolarized $\gam\cpone\cmone$ cross section.
Error estimates for $\mcpmone$ and $\dmchi$ obtained by
studying kinematic distributions of the final pion decay products
were given for two procedures, assuming $L=1\abi$ of accumulated data.
The first employs events in which the $\cpone\cmone$ invariant mass 
is very close to $2\mcpmone$ (i.e. events with very large $p_T^\gamma$).
After accumulating $L=1\abi$ with unpolarized beams,
we find that the location of the 
threshold for the recoil mass opposite the $\gam$ trigger
(i.e. the invariant mass of the $\cpone\cmone$ pair system), denoted
$\mchichi$, gives a very accurate measurement of $\mcpmone$ 
(typically $<0.5\%$ error), and that
the average soft $\pi$ energy, $\epipr$, in the $\mchichi$ center of mass
for $\mchichi$ near threshold gives a very accurate measurement of $\dmchi$
(typically $1\%$). These procedures for detecting $\gam\cpone\cmone$
and measuring $\mcpmone$ and $\dmchi$ using only events
with the highest photon energies, such that $\mchichi\sim 2\mcpmone$,
guarantee that these goals can be achieved even if there is significant
background (contrary to our expectation) at lower $E_\gam$ and $p_T^\gamma$
for the photon tag. Only if $\mcpmone$ is close to $\rts/2$ would
we have to rely on the background being small for $p_T^\gamma$ as
small as 10 GeV. However, if we are correct that
this background remains small for such a low $p_T^\gamma$
for the trigger, slightly better
accuracies for $\mcpmone$ and $\dmchi$
can be achieved (for all, but especially moderate, $\mcpmone$ values) 
by employing the full $[\epipr,\mchichi]$
distribution of the events and applying statistical tests
for discriminating between the distributions obtained for different
$[\mcpmone,\dmchi]$ choices. We emphasize that the low $p_T^\gamma$
events will {\it certainly} be background free if $\dmchi\lsim 600\mev$
(as typical) since in almost all events
one or both of the final $\pi$'s will have an
observable high impact parameter (HIP). 

For the typical case of $\mcpmone=175\gev$ and $\dmchi=200\mev$
(for which essentially all events have a HIP $\pi$
and, hence, backgrounds are certain to be small
even when including all events with $p_T^\gam>10\gev$), we found that 
for $L=1\abi$ the value of $\dmchi$ can be determined to better than 1 MeV 
from the transverse impact parameter distribution of the
final pions. Simultaneously, $c\tau$ for the $\cpmone\to\pi^\pm
\cnone$ decay can be determined to roughly 2\%.
(The impact distribution is less sensitive to $\mcpmone$,
which is best determined from the kinematic distributions.)  

Finally, the underlying parameters $M_2$ and $\mu$ of supersymmetry can be 
most reliably determined from the magnitude of the $\gam\cpone\cmone$
right-handed polarized $e^-$ beam cross section and the 
polarized beam and unpolarized beam angular asymmetries.
As part of the procedure, 
one inputs the precise determination of $\mcpmone$ obtained by
employing the threshold or statistical techniques.
For the typical $\mcpmone=175\gev$, $\dmchi=200\mev$ case, 
we determined the statistical errors for these quantities
assuming $L=1\abi$ each in unpolarized and polarized running.
We then computed the $\Delta\chi^2$ for discriminating
between a given input model choice of $M_2,\mu,\tanb$ (constrained to
give the value of $\mcpmone=175\gev$, presumed to already be well-measured)
and other possible choices consistent 
with the same $\mcpmone$ value.\footnote{Values
of $|\mu|<300\gev$ were not considered; for such values, $\cpmtwo$ production
becomes possible and the strategy for determining SUSY parameters would change
substantially.}
We found the best parameter accuracies
by accumulating luminosity only at the highest energy
($\rts=600\gev$ in our study), distributed roughly equally
between unpolarized beam running and pure $e_R^-$ running.
We found $1\sigma$ (68\% CL) accuracies for $M_2$ and $|\mu|$ of
$\pm 8\%$ and $\pm 16\%$, respectively, for $|\mu|\sim 400\gev$, and
$\pm 8\%$ and $\pm 40\%$, respectively, for $|\mu|\sim 600\gev$.
Errors for both $M_2$ and $|\mu|$ become uselessly large 
by $|\mu|\sim 900\gev$.
The sign of $\mu$ is not determined at the $1\sigma$ level.  However,
the error for $M_2$ decreases to about $5\%$ 
for $350\gev<|\mu|<750\gev$ if the sign of $\mu$ is known
from other input.  Finally, $\tanb$ is essentially undetermined
at the $1\sigma$ level. We have noted that for the small values of
$\dmchi$ natural in the wino LSP scenarios, correlations involving the final
soft $\pi$'s are negligible and do not aid in parameter determination.

In the above, we did not employ the 
absolute normalization of the unpolarized cross section
since it is sensitive to many theoretical uncertainties
at the $(5-10)\%$ level, a level of uncertainty that is larger
than the amount of variation with respect to the parameters
of interest. One source of uncertainty is the unknown $\snue$ mass. 
By measuring the unpolarized $\gam\cpone\cmone$ cross section at
two energies, $m_{\snue}$ could be extracted with an accuracy
determined by the other theoretical uncertainties in $\sigma_{\rm unpolarized}$.
However, if we expend luminosity for this purpose, the errors
for the $M_2$ and $\mu$ determinations would increase.

Of course, additional SUSY signals
{\it will} emerge if some of the squarks, 
sleptons and/or sneutrinos are light enough 
(but still heavier than the $\cpmone$) that 
their production rates are substantial.
In particular, leptonic signals
from the decays [\eg\ $\wtil \ell_L^{\pm}\to \ell^{\pm}\cnone$ or 
$\wtil\nu_{\ell} \to \ell^{\pm} \cmpone$ ] would be present.
Also, depending upon $\rts$ and the mass splitting between the
wino-like $\cnone$ and the bino-like $\cntwo$ (which
is large in many models), the suppressed
$\cnone\cntwo$ production channel might be detectable. Or, the
radiative electroweak symmetry breaking 
scenario chosen by nature could be sufficiently unconventional
that $|\mu|$ is small enough for production of charged and
neutral higgsino-like states to be detectable. We have chosen to emphasize
the case (which is most likely in typical models) that
none of these additional signals are present until $\rts$
substantially above $600\gev$ is available.

It is, of course, likely that the LHC will have been operating for
a number of years prior to the contruction of the next $\epem$ collider.
Because of the large available center of mass energy, it is probable
that some of the heavier states mentioned above will be produced.
However, it is not completely clear that they will be
observed.  The role of the LHC for the type of model being considered
will be discussed in a later paper.

\bigskip

\centerline{\bf Acknowledgements}

This work was supported by the Department of Energy and by the Davis Institute
for High Energy Physics. JFG would like to acknowledge the hospitality of
the Aspen Center for Physics during a portion of this work.

\bigskip
\noindent{\bf 7. Appendix}
\bigskip

In this Appendix we present some details regarding the reconstruction
of the momenta of all the particles for the process
$\epem\to \gam\cpone\cmone\to\gam\pi^+\anti{\cnone}\pi^-\cnone$. (We use
the notation $\anti{\cnone}$ to distinguish the $\cnone$ associated
with the $\cpone$ decay; the $\cnone$ is, of course, its own antiparticle.)
We also give the form of the $\gam\cpone\cmone$ cross section
resulting from the often dominant terms (a gauge invariant subset)
in which the $\gam$ is radiated only from an initial $e^+$ or $e^-$.

The reconstruction of the final state momenta is performed in analogy
with the techniques developed for 
$\epem\to \wp\wm\to \ell^+\nu\ell^-\anti\nu$ \cite{charlton}.
We begin by first using the observed $\gam$ four momentum and
the known $e^+$ and $e^-$ beam momenta to boost to the $\cpone\cmone$
center of mass. The momenta in the expressions below are those
defined in this frame. (We drop the $^*$ superscripts employed in the
main text.) In the $\cpone\cmone$ c.m.s., the 6 components
of the observed $\pi^+$ and $\pi^-$ three momenta, combined
with the known $\mcnone$ and $\mcpmone$ masses, can be used to solve
for the three momenta of the $\cnone$ and $\anti{\cnone}$
up to a two-fold ambiguity indicated by the $\pm d$ term below:
\beq
\pn=a\km+b\kp \pm d (\crs)\,,\quad \pnp=-(\pn+\kp+\km)\,,
\label{prec}
\eeq
where $a$, $b$ and $d$ are given by:\footnote{We correct some errors in
the formulae of Ref.~\cite{charlton}.}
\bea
a&=& {m|\kp|^2-n\dt\over |\crs|^2}\,,\quad b={-m\dt+n|\km|^2\over |\crs|^2}\,,
\nonumber\\
d^2&=&{(E_b-\Em)^2-\mcnone^2-a^2|\km|^2-b^2|\kp|^2-2ab\dt\over |\crs|^2}\,,
\label{abd}
\eea
where 
\beq
m\equiv -{1\over 2}\left(\mcpmone^2-\mcnone^2-2E_b\Em+\Em^2+|\km|^2\right)\,,
\quad
n\equiv {1\over 2}\left(\mcpmone^2-\mcnone^2-2E_b\Ep-2\dt\right)\,.
\label{mndef}
\eeq
Using these formulae, we can, up to the two-fold ambiguity,
reconstruct the $\cpone$ and $\cmone$ momenta.
The construction only fails if $\crs=0$ or if $d^2<0$ (the latter being
possible as a result of momentum
smearing in the detector). If $d^2<0$, 
setting $d=0$ usually gives a fairly accurate result
for $\pn$ and $\pnp$. It is perhaps useful to keep in mind approximations
that follow from the fact that $\Ep,\Em,|\kp|,|\km|\sim \dmchi$ and 
$\mcpmone^2-\mcnone^2\sim 2\mcnone\dmchi$.
For small $\dmchi/\mcnone$ we have
\beq 
m=E_b\Em-\mcnone\dmchi\,,\quad n=-E_b\Ep+\mcnone\dmchi\,,
\quad d={\sqrt{E_b^2-\mcnone^2}\over |\crs|}\,.
\label{approxs}
\eeq
Further, $\pcp=\pnp+\km\sim \pnp$ and $\pc=\pn+\kp\sim \pn$. 
Finally, we note that if $E_b>\mcpmone\dmchi/\mpi$, the $\pi$ directions are 
guaranteed to
have positive dot product with the directions of their parent charginos.

The utility of this reconstruction follows from an understanding
of how the matrix element squared depends upon the final state momenta.
As described in the text, correlations between the helicity of the $\cpmone$
and the three-momentum of the decay pion are of order $\dmchi/\mcpmone$.
Thus, to a very good approximation
the distribution of the decay pion 
in the $\cpmone$ rest frame is completely independent of the
$\theta^*,\phi^*$ rest frame decay angles for any fixed helicity
of the $\cpmone$. The final state only contains 
information encoded in
the decay distributions of the $\cpone,\cmone$.
Thus, aside from small terms of order $\dmchi/\mcpmone$,
the decay pions do not have any correlations
other than those kinematically induced by the boosts in
the directions of their parent charginos and
the form of the cross section as a function of $\theta$, 
the angle of the $\cmone$ in the $\cpone$-$\cmone$ center-of-mass.

For the case of a photon radiated from the initial $e^+$ or $e^-$, the form
of the invariant matrix element squared for $\epem\to\gam\cpone\cmone$ is
$|{\cal A}|^2=\left(2\pep\cdot\pg\,\pem\cdot\pg\right)^{-1}|{\cal B}|^2$ with
\bea
|{\cal B}|^2 &\propto& 
 - 2 \pc\cdot\pem \pcp\cdot\pem \pep\cdot\pg 
(\qll^2 + \qlr^2 + \qrl^2 + \qrr^2) 
\nonumber\cr
\phantom{|{\cal A}|^2} &\phantom{\propto}&
- 2 \pc\cdot
\pem \pcp\cdot\pep s\, (\qlr^2 + \qrl^2) 
\nonumber\cr
\phantom{|{\cal A}|^2} &\phantom{\propto}&
+ 2 \pc\cdot\pem \pcp\cdot\pep (\pem\cdot\pg \qlr^2 + 
\pem\cdot\pg \qrl^2 + \pep\cdot\pg \qlr^2 + \pep\cdot\pg \qrl^2) 
\nonumber\cr
\phantom{|{\cal A}|^2} &\phantom{\propto}&
+ \pc\cdot\pem s\, \pcp\cdot\pg (\qlr^2 + \qrl^2) 
\nonumber\cr
\phantom{|{\cal A}|^2} &\phantom{\propto}&
- 2 \pc\cdot\pem \pcp\cdot\pg \pem\cdot\pg (\qlr^2 + \qrl^2) 
\nonumber\cr
\phantom{|{\cal A}|^2} &\phantom{\propto}&
- 2 \pc\cdot\pep \pcp\cdot\pem s\, (\qll^2 + \qrr^2) 
\nonumber\cr
\phantom{|{\cal A}|^2} &\phantom{\propto}&
+ 2 \pc\cdot\pep \pcp\cdot\pem (\pem\cdot\pg \qll^2 + 
\pem\cdot\pg \qrr^2 + \pep\cdot\pg \qll^2 + \pep\cdot\pg \qrr^2) 
\nonumber\cr
\phantom{|{\cal A}|^2} &\phantom{\propto}&
- 2 \pc\cdot\pep \pcp\cdot\pep \pem\cdot\pg 
(\qll^2 + \qlr^2 + \qrl^2 + \qrr^2) 
\nonumber\cr
\phantom{|{\cal A}|^2} &\phantom{\propto}&
+ \pc\cdot\pep s\, \pcp\cdot\pg (\qll^2 + \qrr^2) 
\nonumber\cr
\phantom{|{\cal A}|^2} &\phantom{\propto}&
- 2 \pc\cdot\pep \pcp\cdot\pg \pep\cdot\pg (\qll^2 + \qrr^2) 
\nonumber\cr
\phantom{|{\cal A}|^2} &\phantom{\propto}&
+ \pcp\cdot\pem s\, \pc\cdot\pg (\qll^2 + \qrr^2) 
\nonumber\cr
\phantom{|{\cal A}|^2} &\phantom{\propto}&
- 2 \pcp\cdot\pem \pc\cdot\pg \pem\cdot\pg (\qll^2 + \qrr^2) 
\nonumber\cr
\phantom{|{\cal A}|^2} &\phantom{\propto}&
+ \pcp\cdot\pep s\, \pc\cdot\pg (\qlr^2 + \qrl^2) 
\nonumber\cr
\phantom{|{\cal A}|^2} &\phantom{\propto}&
- 2 \pcp\cdot\pep \pc\cdot\pg \pep\cdot\pg (\qlr^2 + \qrl^2) 
\nonumber\cr
\phantom{|{\cal A}|^2} &\phantom{\propto}&
- s^2 \mcpmone^2 (\qll \qlr + \qrl \qrr) 
\nonumber\cr
\phantom{|{\cal A}|^2} &\phantom{\propto}&
+ 2 s\, \mcpmone^2 (\pem\cdot\pg \qll \qlr + 
\pem\cdot\pg \qrl \qrr + \pep\cdot\pg \qll \qlr + \pep\cdot\pg \qrl \qrr) 
\nonumber\cr
\phantom{|{\cal A}|^2} &\phantom{\propto}&
- 2 \mcpmone^2 ((\pem\cdot\pg)^2 \qll \qlr + (\pem\cdot\pg)^2 \qrl \qrr + 
(\pep\cdot\pg)^2 \qll \qlr + (\pep\cdot\pg)^2 \qrl  \qrr) \,.
\label{ampsqgam}
\eea
In the above, all $Q_{\alpha\beta}$'s are evaluated at $\mchichi^2$.
Note that $|{\cal A}|^2$
reduces to Eq.~(\ref{ampsq}) in the $p_\gam\to 0$ limit after removing
the photon radiation pole factors $(p_{e^+}\cdot p_\gam)(p_{e^-}\cdot p_\gam)$.
This answer is gauge invariant on its own.
Contributions to $|{\cal A}|^2$ coming from 
diagrams in which the photon
is radiated from one of the final $\cpmone$ lines are often
suppressed relative to the initial state radiation terms given above.
All diagrams were included in our numerical calculations.
In practice, there are delicate cancellations between large terms in
Eq.~(\ref{ampsqgam}), and also in the full expression for $|{\cal A}|^2$
obtained after including all diagrams, that can lead to numerical 
inaccuracies.  Thus, it is actually best to sum all the
diagrams numerically at the  matrix element level
and then square.  This is the approach used in our numerical simulations.



\begin{thebibliography}{99}
\bibitem{charlton} D. Charlton, OPAL Techical Note TN536. 
Aside from changes necessitated by $\mpi\neq0$ and $\mcnone\neq0$, several
typos and one inadvertent error in the Appendix of this paper are corrected.
\end{thebibliography}
\end{document}